\documentclass[sigconf, screen]{acmart}

\usepackage{colortbl}
\usepackage{multirow}
\usepackage{graphicx}
\usepackage{subcaption}
\usepackage{cleveref}
\usepackage{stfloats}
\usepackage{tcolorbox}
\usepackage{array}
\usepackage{adjustbox}
\usepackage{booktabs}
\usepackage{graphicx}
\tcbuselibrary{most}
\usepackage{listings} 
\usepackage{xcolor}
\usepackage{xspace}
\usepackage{color}
\usepackage{url}
\usepackage{listings}
\usepackage{ifthen}
\usepackage{enumitem}
\usepackage{makecell}
\renewcommand\theadfont{\scriptsize}

\usepackage{amssymb}
\usepackage{float}
\usepackage{threeparttable}
\usepackage{diagbox}
\usepackage{amsmath}
\usepackage{lipsum}
\usepackage[ruled,vlined]{algorithm2e}
\usepackage{algorithmic}
\usepackage{colortbl}

\SetCommentSty{mycommfont}
\AtBeginDocument{%
	\providecommand\BibTeX{{%
			\normalfont B\kern-0.5em{\scshape i\kern-0.25em b}\kern-0.8em\TeX}}}

\newcommand{\todoteal}[1]{\textcolor{teal}{~#1}}
\newcommand{\todoorange}[1]{\textcolor{orange}{~#1}}
\definecolor{mygreen}{HTML}{238b45}
\newcommand{\todogreen}[1]{\textcolor{mygreen}{~#1}}
\newcommand{\todoviolet}[1]{\textcolor{violet}{~#1}}
\newcommand{\todoblue}[1]{\textcolor{blue}{~#1}}
\newcommand{\todopink}[1]{\textcolor{pink}{~#1}}

\newcommand{\keepnotes}{true}

\ifthenelse{\isundefined{\keepnotes}}{
\newcommand{\civi}[1]{}
\newcommand{\zichao}[1]{}
\newcommand{\bella}[1]{}
\newcommand{\scc}[1]{}
\newcommand{\xiaohu}[1]{}
\newcommand{\shangwen}[1]{}
}{
\newcommand{\scc}[1]{\todoblue{[scc: #1]}}
\newcommand{\zichao}[1]{\todoteal{[zichao: #1]}}
\newcommand{\civi}[1]{\todogreen{[civi: #1]}}
\newcommand{\xiaohu}[1]{\todoorange{[xiaohu: #1]}}
\newcommand{\bella}[1]{\todoviolet{[bella: #1]}}
\newcommand{\shangwen}[1]{\todopink{[shangwen: #1]}}
}
\newcommand*{\myfont}{\fontfamily{LinuxBiolinumT-OsF}\selectfont}
\newcommand*{\eg}{e.g., }
\newcommand*{\mycode}{\fontfamily{lmtt}\selectfont}

\newcommand*{\ie}{i.e., }

\newcommand{\etal}{\emph{et~al.}\xspace}

\newcommand{\tabincell}[2]
{\begin{tabular}{@{}#1@{}}#2\end{tabular}}

\setcopyright{acmlicensed}
\acmPrice{15.00}
\acmDOI{10.1145/3611643.3616356}
\acmYear{2023}
\copyrightyear{2023}
\acmSubmissionID{fse23main-p1385-p}
\acmISBN{979-8-4007-0327-0/23/12}
\acmConference[ESEC/FSE '23]{Proceedings of the 31st ACM Joint European Software Engineering Conference and Symposium on the Foundations of Software Engineering}{December 3--9, 2023}{San Francisco, CA, USA}
\acmBooktitle{Proceedings of the 31st ACM Joint European Software Engineering Conference and Symposium on the Foundations of Software Engineering (ESEC/FSE '23), December 3--9, 2023, San Francisco, CA, USA}
\received{2023-03-02}
\received[accepted]{2023-07-27}

\begin{document}
	\title{An Extensive Study on Adversarial Attack against Pre-trained Models of Code}
	
	\author{Xiaohu Du}
	\affiliation{%
		\institution{Huazhong University of Science\\ and Technology, China}
            \country{}
        }
	\email{xhdu@hust.edu.cn}
        \authornote{National Engineering Research Center for Big Data Technology and System, Services Computing Technology and System Lab, HUST, Wuhan, 430074, China}
        \authornote{Hubei Engineering Research Center on Big Data Security, Hubei Key Laboratory of Distributed System Security, School of Cyber Science and Engineering, HUST, Wuhan, 430074, China}

        \author{Ming Wen}
	\affiliation{%
		\institution{Huazhong University of Science\\ and Technology, China}
            \country{}
        }
	\email{mwenaa@hust.edu.cn}
        \authornotemark[1]
        \authornotemark[2]
        \authornote{Corresponding author}

        \author{Zichao Wei}
	\affiliation{%
		\institution{Huazhong University of Science\\ and Technology, China}
            \country{}
        }
	\email{u201911736@hust.edu.cn}
        \authornotemark[1]
        \authornotemark[2]

        \author{Shangwen Wang}
	\affiliation{%
		\institution{National University of Defense Technology, China}
            \country{}
        }
	\email{wangshangwen13@nudt.edu.cn}

        \author{Hai Jin}
	\affiliation{%
		\institution{Huazhong University of Science\\ and Technology, China}
            \country{}
        }
	\email{hjin@hust.edu.cn}
        \authornotemark[1]
        \authornote{Cluster and Grid Computing Lab, School of Computer Science and Technology, HUST, Wuhan, 430074, China} 
	
	\renewcommand{\shortauthors}{Xiaohu Du, Ming Wen, Zichao Wei, Shangwen Wang, and Hai Jin}
	\begin{abstract}
Transformer-based \textit{pre-trained models of code} (PTMC) have been widely utilized and have achieved state-of-the-art performance in many mission-critical applications. However, they can be vulnerable to adversarial attacks through identifier substitution or coding style transformation, which can significantly degrade accuracy and may further incur security concerns.
Although several approaches have been proposed to generate adversarial examples for PTMC, the effectiveness and efficiency of such approaches, especially on different code intelligence tasks, has not been well understood. 
To bridge this gap, this study systematically analyzes five state-of-the-art adversarial attack approaches from three perspectives: effectiveness, efficiency, and the quality of generated examples.
The results show that none of the five approaches balances all these perspectives. 
Particularly, approaches with a high attack success rate tend to be time-consuming; the adversarial code they generate often lack naturalness, and vice versa.
To address this limitation, we explore the impact of perturbing identifiers under different contexts and find that identifier substitution within {\mycode for} and {\mycode if} statements is the most effective. Based on these findings, we propose a new approach that prioritizes different types of statements for various tasks and further utilizes  beam search to generate adversarial examples. 
Evaluation results show that it outperforms the state-of-the-art ALERT in terms of both effectiveness and efficiency while preserving the naturalness of the generated adversarial examples.
\end{abstract}

	
	
	\maketitle
	
	\vspace{-2mm}
\section{Introduction}
Given the rapid development of \textit{deep learning} (DL), many researchers have applied DL techniques in various programming language processing tasks with promising results achieved over recent years, and such a trend continuously rises. 
The recently-proposed Transformer architecture~\cite{DBLP:conf/nips/VaswaniSPUJGKP17}, which mainly employs the self-attention mechanism, has shown promising results on dealing with the long range dependency problem, which is a critical challenge for traditional sequence models such as the Recurrent Neural Network.
Therefore, a number of state-of-the-art DL models are designed based on such an architecture, one category of  which is the \textit{pre-trained models of code} (PTMCs), such as CodeBERT~\cite{DBLP:conf/emnlp/FengGTDFGS0LJZ20} and  CodeGPT ~\cite{DBLP:conf/nips/LuGRHSBCDJTLZSZ21}.

Via utilizing the pre-training techniques, domain knowledge in the large-scale publicly-available code repositories can be learned by PTMCs, which can be further leveraged on downstream tasks such as vulnerability detection, clone detection, and code summarization~\cite{DBLP:conf/emnlp/FengGTDFGS0LJZ20, DBLP:conf/nips/LuGRHSBCDJTLZSZ21, DBLP:conf/naacl/AhmadCRC21}.
Unfortunately, recent studies have shown that similar to conventional deep learning models in the domains of computer vision and natural language processing, PTMCs can also generate totally different results given two semantically-identical input programs, one of which (a.k.a. the {\em adversarial example}) is generated by performing certain semantic-preserving transformations to the other \cite{DBLP:conf/aaai/ZhangLLMLJ20, DBLP:journals/tosem/ZhouZSHCG22, DBLP:conf/icse/YangSH022, DBLP:conf/issta/ZengTZLZZ22, DBLP:conf/icse/LiCCZX22, DBLP:journals/infsof/RabinBWYJA21, DBLP:conf/wcre/HenkelRWAJR22, DBLP:conf/icst/PourL0H21}.
This is devastating considering that PTMCs have been deployed to a wide range of mission-critical applications such as vulnerability detection~\cite{DBLP:conf/icse/YangSH022,DBLP:journals/tosem/ZhangFLMZYSLJ22}. 
Specifically, an attacker may easily generate an adversarial example that retains the vulnerability while misleading the PTMC to label it as ``non-vulnerable''. 

One potential way to alleviate such a threat is adversarial re-training, where the models under attack are continuously trained with the generated adversarial examples to enhance the robustness~\cite{DBLP:conf/icse/YangSH022, DBLP:journals/pacmpl/Yefet0Y20}.
Therefore, over the years, a number of adversarial attack approaches have been proposed aiming to automatically generate adversarial examples~\cite{DBLP:conf/aaai/ZhangLLMLJ20, DBLP:journals/tosem/ZhouZSHCG22, DBLP:conf/icse/YangSH022, DBLP:conf/issta/ZengTZLZZ22, DBLP:conf/icse/LiCCZX22, DBLP:journals/infsof/RabinBWYJA21, DBLP:conf/wcre/HenkelRWAJR22, DBLP:conf/icst/PourL0H21}.
Existing adversarial attack approaches differ in terms of various design aspects. First, at a high level, the semantic-preserving transformation can be performed at both the \textbf{token level} (\eg by identifier substitution~\cite{DBLP:conf/icst/PourL0H21}) and the \textbf{statement level} (\eg by adding dead code~\cite{DBLP:journals/pacmpl/Yefet0Y20}).
Second, even if certain approaches are designed at the same granularity (\eg~identifier substitution), there also exists multiple choices when determining how and what identifiers to be replaced: by random selection~\cite{DBLP:conf/issta/ZengTZLZZ22} or via pre-defined heuristics~\cite{DBLP:conf/icse/YangSH022}. 
Such design aspects may significantly affect the effectiveness of the attack approaches. For instance, Yang \etal~\cite{DBLP:conf/icse/YangSH022} showed that certain pre-defined heuristics could outperform random selection on generating new identifier names.

Although huge efforts have been made towards advancing adversarial attacks targeting for PTMCs, the performance of existing techniques has not been systematically evaluated and compared. Little is known about their advantages and disadvantages. 
There is thus an urgent need for a comprehensive empirical study comparing and analyzing the effectiveness of the \textit{state-of-the-art} (SOTA) adversarial attacks targeting PTMCs. 
In particular, how is the effectiveness and efficiency of the SOTA approaches with respect to various PTMCs? 
How well do different approaches generalize across various code intelligence tasks? Most importantly, how is the quality of the adversarial examples generated by different approaches? 
If the quality is extremely low, the practical usefulness can be compromised since they can be easily perceived by developers.  
Additionally, it is unclear how the context of code perturbations affects the effectiveness of current attack approaches. Understanding such problems is important to guide future researches in this field.

To fill this gap, in this study, we perform an extensive study on existing SOTA adversarial attack approaches against PTMCs. Specifically, we utilize five SOTA adversarial attack approaches to attack three widely-used PTMCs (\eg CodeBERT~\cite{DBLP:conf/emnlp/FengGTDFGS0LJZ20}, CodeGPT~\cite{DBLP:conf/nips/LuGRHSBCDJTLZSZ21}, and PLBART~\cite{DBLP:conf/naacl/AhmadCRC21}). 
Our evaluation is performed on three well-studied code intelligence tasks, including one generation task (\ie code summarization) and two understanding tasks (\ie vulnerability detection and code clone detection).
Through extensive evaluations and comparisons, our study makes several interesting findings: (1) PTMCs can be easily attacked under all the three tasks and they are relatively less robust on the generation tasks compared with understanding tasks;
(2) There is a trade-off between the effectiveness and efficiency for the adversarial attacks: the attack approach with the highest success rate usually queries PTMCs for the most times;
(3) The quality of adversarial examples is heavily influenced by the identifier substitution strategy. Identifiers predicted with context-aware information produce the highest quality examples that are very similar to the original code, followed by a cosine similarity-based substitution strategy. On the other hand, random substitution leads to the lowest quality adversarial examples;
and (4) replacing identifiers under different types of statements exhibits diverse chances to generate adversarial examples successfully while such chances differ significantly with respect to the generation and understanding tasks.

Based on our findings, we design an efficient yet effective attack approach called BeamAttack for code adversarial attack.
BeamAttack separates identifiers into several groups based on the statements where they are extracted. It then iteratively selects identifiers in a prioritized manner, selecting those that are most likely to result in successful attacks, as summarized by our empirical evaluation.
BeamAttack reduces the attacking costs by dividing identifiers into smaller sub-groups and prioritizing them based on the likelihood of successful attacks, rather than searching the entire identifier space. It can also reduce the risk of getting stuck in local optima, as opposed to searching each individual identifier similar to WIR~\cite{DBLP:conf/issta/ZengTZLZZ22}.
The results on a total of six datasets demonstrate
that our approach achieves higher attack success rates with less queries than ALERT while can preserve the naturalness of the generated adversarial examples (\ie the generated examples bear a high resemblance to the code written by humans).

To summarize, we make the following major contributions:
\vspace{-1mm}
\begin{itemize}[leftmargin=*]
\item \textbf{Originality.} To our best knowledge, we perform the first extensive study on existing SOTA adversarial attacks approaches towards PTMCs under well-studied code intelligence tasks. 
 \item \textbf{Extensive Study.} We systematically compare five state-of-the-art adversarial attack approaches from three perspectives: effectiveness, efficiency, and the quality of generated examples.
 Our evaluation reveals the strengths and weaknesses of existing approaches, highlights useful insights, thus paving the way for future researches in this field.
 \item \textbf{Improvement.} Based on our empirical findings, we exploit the differences among diverse program contexts with respect to the chances of successfully generating adversarial examples and design a simple yet effective attack approach. Our approach has demonstrated promising results via extensive evaluation. 
 \item \textbf{Open Science.} We have released all the artefacts of our study, including the source code and experiment results, which available at: \url{https://github.com/CGCL-codes/Attack_PTMC}. 
    
\end{itemize}

	\vspace{-2mm}
\section{Background}
\subsection{Pre-Trained Models of Code}\label{ptmc}
PTMC can learn universal language representations on the large corpus and can avoid training a new model from scratch~\cite{DBLP:journals/corr/abs-2003-08271, DBLP:journals/aiopen/HanZDGLHQYZZHHJ21}. 
The pre-training paradigm usually consists of two stages: pre-training and fine-tuning. In the pre-training stage, it captures generic language knowledge by employing self-supervised learning on a large unlabeled corpus. In the fine-tuning stage, the trained model can be fine-tuned for different downstream tasks. 
PTMC can be divided into three categories based on their architectures: encoder-only, decoder-only, and encoder-decoder models~\cite{DBLP:conf/issta/ZengTZLZZ22}. 
Encoder-only pre-trained models can support both the understanding and generation tasks, and the most widely used ones are CodeBERT~\cite{DBLP:conf/emnlp/FengGTDFGS0LJZ20} and GraphCodeBERT~\cite{DBLP:conf/iclr/GuoRLFT0ZDSFTDC21}. 
Decoder-only models are good at generation tasks like code completion while the adopted  unidirectional architectures are less effective on understanding tasks such as clone detection~\cite{DBLP:conf/acl/GuoLDW0022}.
CodeGPT~\cite{DBLP:conf/nips/LuGRHSBCDJTLZSZ21} is a well-known model based on Transformer belonging to this category. 
Encoder-decoder models are proposed aiming to tackle both the understanding and generation tasks, and PLBART~\cite{DBLP:conf/naacl/AhmadCRC21} as well as CodeT5~\cite{DBLP:conf/emnlp/0034WJH21} are typical ones of such models. 

\vspace{-2mm}
\subsection{Adversarial Attack on Code}\label{definitions}
\subsubsection{Code Processing Tasks}\label{sec:task}
Following prior works~\cite{DBLP:conf/issta/ZengTZLZZ22, DBLP:conf/nips/LuGRHSBCDJTLZSZ21}, we briefly introduce three typical code processing tasks, which involve the understanding task and generation task.

\noindent\textbf{Clone Detection.}
It is a program understanding task aiming to detect whether two source code snippets are identical or similar. 

\noindent\textbf{Vulnerability Detection.} 
It is another program understanding task whose purpose is to determine if a given code snippet contains vulnerabilities or not.

\noindent\textbf{Code Summarization.}
It is a generation task, which aims to generate natural language texts that describe the functionality of a given code snippet.

\vspace{-2mm}
\subsubsection{Definitions}
We give two definitions respectively for the understanding task and generation task. 
For the understanding task, a classifier $f: \mathcal{X} \rightarrow \mathcal{Y}$ is expected to predict the ground-truth label $y_{truth} \in \mathcal{Y}$ for a given code snippet $x \in \mathcal{X}$.
The goal of adversarial attack is to add slight perturbations on $x$ to generate adversarial examples $x^{adv}$ that can mislead $f$.
Specifically, an adversarial code example should satisfy the following three requirements: (1) Adversarial example should mislead the model: $f(x^{adv})\neq f(x) = y_{truth}$. (2)  Adversarial perturbations should ensure the code is still syntactically correct. That is, perturbations should conform to the syntax rules of the programming language. 
For example, for the C language, the identifiers can only contain letters, numbers, and underscores.
(3) $x^{adv}$ should be semantically equivalent to $x$ (\ie~have exactly the same functionalities and generate the same results given a same input).
For the generation task, we take the code summarization as an example.
The model $f: \mathcal{X} \rightarrow \mathcal{Y}$ aims to maximize $P\left(y_{truth} \mid x\right)$ where a given code snippet $x \in \mathcal{X}$ and the ground-truth summary $y_{truth} \in \mathcal{Y}$.
Since the output $y$ of summarization models contains many possibilities, we cannot employ the first requirement in the understanding task to directly determine whether the attack is successful.
The existing work~\cite{DBLP:journals/tosem/ZhouZSHCG22} utilizes the decrease on the BLEU score to evaluate the performance of attack approaches. 
In this paper, we follow the existing study \cite{DBLP:conf/issta/ZengTZLZZ22} to consider an attack successful when the BLEU score between adversarial summary and the reference summary is 0, which indicates that the adversarial summary does not match the reference summary at all.
Similarly, adversarial examples in generation tasks should also meet the requirements (2) and (3) as defined in the understanding tasks.

	\vspace{-2mm}
\section{Study Design}

In this section, we introduce the design of our empirical study, including the selected pre-trained models, adversarial attack approaches, and benchmark datasets. 
We then introduce our designed research questions and the corresponding experimental settings. 

\vspace{-2mm}
\subsection{Subjects and Datasets}

\subsubsection{Target Models}\label{sec:model}
Section~\ref{ptmc} presents the SOTA PTMCs to date. 
For each category of PTMCs, we choose one model for evaluation as the previous report~\cite{DBLP:conf/issta/ZengTZLZZ22} indicates that they achieve very close performance.
Meanwhile, there is no PTMC model that can achieve the optimum performance across different tasks and datasets (\eg~for encoder-only model, CodeBERT is better than GraphCodeBERT in vulnerability detection while vice versa in clone detection. Similarly, for the encoder-decoder model, CodeT5 outperforms PLBART in vulnerability detection while vice versa in clone detection \cite{DBLP:conf/issta/ZengTZLZZ22}).
Therefore, we select the most popular and widely-used one (indicated by the number of citations) for each category.
In particular, we select CodeBERT~\cite{DBLP:conf/emnlp/FengGTDFGS0LJZ20}, CodeGPT~\cite{DBLP:conf/nips/LuGRHSBCDJTLZSZ21}, and PLBART~\cite{DBLP:conf/naacl/AhmadCRC21} in our study from each category. 

\vspace{-2mm}
\subsubsection{Adversarial Attack Approaches}\label{sec:attack_method}
Table~\ref{Summary} summarizes the state-of-the-art adversarial attack approaches published in the major conferences and journals. 
The approaches selected in this study are all black-box approaches since
(1) white-box attacks often require to access the information of model structures and parameters which might not be easily obtained in practice, and thus attackers typically can only access the provided APIs to query the model; and 
(2) white-box attacks tend to be model-specific in that different models employ different structures, and thus an attack approach against a specific model cannot generalize well to other ones.
However, in this study, we aim to evaluate the selected attack approach against different models under various applications.
Among the listed black-box attacks, we exclude the approach proposed by Nguyen et al.~\cite{DBLP:conf/kbse/NguyenSRPR21} because it performs fake API insertion at the class level while the datasets selected to evaluate the three tasks in this study are all at the function level.
Among the remaining eight black-box attacks,
four different attacks~\cite{DBLP:conf/aaai/ZhangLLMLJ20, DBLP:journals/tosem/ZhouZSHCG22, DBLP:conf/icse/YangSH022, DBLP:conf/issta/ZengTZLZZ22} can be directly reproduced and are thus selected as our study subjects.
As for the remaining four approaches~\cite{DBLP:conf/icse/LiCCZX22, DBLP:journals/infsof/RabinBWYJA21, DBLP:conf/wcre/HenkelRWAJR22, DBLP:conf/icst/PourL0H21}, they perform semantic-preserving transformations at the statement level (\eg by inserting dead code or transforming {\mycode for} loop into {\mycode while} loop).
Because such approaches usually contain common and similar transformation strategies, we summarize widely-used strategies and integrate them into one approach.
We briefly introduce the selected five approaches below.

\noindent\textbf{MHM}~\cite{DBLP:conf/aaai/ZhangLLMLJ20}. MHM performs iterative identifier substitution based on \textit{Metropolis-Hastings} (M-H) sampling~\cite{metropolis1953equation}.
This attack has two main hyperparameters, the maximum number of iterations and the number of candidate identifiers per iteration. 
The larger the value, the higher chance the attack will be successful. 
Unfortunately, it will be less efficient at the same time. 
We set these two parameters to 100 and 30 respectively, following the original paper~\cite{DBLP:conf/aaai/ZhangLLMLJ20}.

\noindent\textbf{ACCENT}~\cite{DBLP:journals/tosem/ZhouZSHCG22}. ACCENT first selects K candidates for each identifier based on the cosine distance, and then selects the best identifier and candidate based on the change in scores before and after substituting the identifiers. 
This approach has two main hyperparameters, the number of candidate identifiers \textit{K} and the number of the identifiers \textit{max} that can be replaced.
For a fair comparison with other attacks, we set k to 30, and cancel the parameter \textit{max} which means we do not limit the number of replaced identifiers.

\noindent\textbf{WIR-Random}~\cite{DBLP:conf/issta/ZengTZLZZ22}. WIR-Random utilizes \textit{Word Importance Rank} (WIR) to determine the substitution sequence of identifiers, which ranks each identifier according to the difference in the probabilities generated by the model before and after renaming the identifier to ``UNK''.
Then, WIR-Random sequentially replaces the sorted identifiers by randomly selecting candidates.
For fair comparison with MHM, we also limit the number of candidate identifiers to 30.

\noindent\textbf{ALERT}~\cite{DBLP:conf/icse/YangSH022}. ALERT utilizes context-aware identifier prediction for substitution. 
In particular, in terms of the identifier selection strategy, ALERT adopts two methods, the greedy algorithm and the genetic algorithm.
We set the relevant hyperparameters following the original paper, including the number of candidate identifiers (\ie 30) and the maximum number of iterations (the larger one between $5 \times Num_{i}$ and 10, where $Num_{i}$ denotes the number of identifiers in the code). 

\noindent\textbf{StyleTransfer}~\cite{DBLP:conf/icse/LiCCZX22, DBLP:journals/infsof/RabinBWYJA21, DBLP:conf/wcre/HenkelRWAJR22, DBLP:conf/icst/PourL0H21}. {The idea of StyleTransfer is to perform certain transformations that do not alter the semantics of the program.}
In this attack, we select some common transformation strategies from existing studies including (1) randomly adding a {\mycode log} statement; (2) replacing {\mycode while} and {\mycode for} loops with each other; (3) exchanging two independent statements; (4) reordering a binary condition; (5) exchanging {\mycode switch} to {\mycode if}; (6) randomly adding a {\mycode try-catch} block; (7) randomly adding a piece of dead code; (8) switching the value of a {\mycode boolean} variable and propagating this change.
Then, we apply transformations to generate \textit{N}  candidate examples and use them to attack the model.
\textit{N} is set to 500 in this study to avoid the huge overhead in the attack process. 
\begin{table}[t]
    \scriptsize
    \caption{Summary of existing adversarial attack approaches (in ascending order by the publication year)}
    \label{Summary}
    \renewcommand\arraystretch{1.}
    \setlength\tabcolsep{1.6pt}
    \centering
    \vspace{-4mm}
    \begin{tabular}{l|c|c|c|c}
        \toprule
           Approach  & Venue & \thead{White/ \\ Black} & Task & Perturbation \\
        \midrule
        \tabincell{c}{DAMP~\cite{DBLP:journals/pacmpl/Yefet0Y20} } &  OOPSLA'20                         & White     &\tabincell{c}{Functionality Classification \\ Code Completion}
        &\tabincell{l}{Random Substitution \\ Dead-Code Insertion }
                 \\  \hline
        \tabincell{l}{MHM~\cite{DBLP:journals/tosem/ZhouZSHCG22}}  & AAAI'20     & Black     &Functionality Classification 
            & Random Substitution \\  \hline
        \tabincell{l}{Srikant~\emph{et al.}~\cite{DBLP:conf/iclr/Srikant0MCFZO21}} & ICLR'21 & White &Functionality Classification
                &\tabincell{c}{Random Substitution \\ Dead-Code Insertion}  \\  \hline
        \tabincell{l}{Rabin~\emph{et al.}~\cite{DBLP:journals/infsof/RabinBWYJA21}} & IST'21 & \tabincell{c}{White \&\\ Black}  &Method Name Prediction
                &\tabincell{c}{Random Substitution \\ Style Transformation}  \\  \hline
        \tabincell{l}{Pour~\emph{et al.}~\cite{DBLP:conf/icst/PourL0H21} } & ICST'21 & Black &\tabincell{c}{Method Name Prediction \\ Code Captioning \\ Code Search \\ Code Summarization} & \tabincell{c}{Random Substitution \\ Style Transformation}  \\  \hline
        \tabincell{l}{Nguyen~\emph{et al.}~\cite{DBLP:conf/kbse/NguyenSRPR21}} & ASE'21 & Black & API Recommender & Fake APIs Insertion \\  \hline
        \tabincell{l}{AVERLOC~\cite{DBLP:conf/wcre/HenkelRWAJR22}} & SANER'22 & Black & Code Summarization & \tabincell{c}{Random Substitution \\ Style Transformation}  \\  \hline
        \tabincell{l}{ACCENT~\cite{DBLP:journals/tosem/ZhouZSHCG22}} & TOSEM'22        & Black     
            &Code Summarization          & Based on Cosine Distance  \\  \hline
        \tabincell{l}{ALERT~\cite{DBLP:conf/icse/YangSH022}} & ICSE'22       & Black     
             &\tabincell{c}{Authorship Attribution \\  
                                            Clone Detection \\                                      Vulnerability Detection} 
                & Context Prediction    \\  \hline
        \tabincell{l}{RoPGen~\cite{DBLP:conf/icse/LiCCZX22}} & ICSE'22           & Black     
                &Authorship Attribution       
                & \thead{Style Transformation} \\  \hline
        \tabincell{l}{CARROT~\cite{DBLP:journals/tosem/ZhangFLMZYSLJ22} } & TOSEM'22        & White     
                &\tabincell{c}{Functionality Classification  \\ Clone Detection \\ Vulnerability Detection}
                & Random Substitution \\  \hline
        \tabincell{l}{WIR-Random~\cite{DBLP:conf/issta/ZengTZLZZ22}} & ISSTA'22 & Black &\tabincell{c}{Vulnerability Detection \\ Code Summarization} & Random Substitution \\
        \bottomrule                                              
    \end{tabular}
    \vspace{-6mm}
\end{table}

\subsubsection{Datasets}\label{sec:dataset}
To ensure the comprehensiveness of our understanding towards the performance of existing attacks, 
we study three tasks: \textit{vulnerability detection}, \textit{clone detection} for understanding tasks, and \textit{code summarization} for generation tasks.
We select representative benchmarks to evaluate them.
For clone detection, BigCloneBench~\cite{DBLP:conf/icsm/SvajlenkoIKRM14} is a widely used clone detection benchmark that contains four main types of intra-project and inter-project clones. 
To better evaluate the adversarial attacks, we adopt the filtered dataset proposed by Yang ~\emph{et~al.}~\cite{DBLP:conf/icse/YangSH022}. Their filtering strategies include removing unlabeled data, balancing the two labels (clones and non-clones), and making the data at a computationally friendly scale. 
As a result, our dataset includes 90,102 examples for training and 4,000 examples for validation and testing, respectively.
For vulnerability detection, the \textit{Open Web Application Security Project} (OWASP) Benchmark\footnote{\url{https://owasp.org/www-project-benchmark}} is a Java test suite designed to evaluate vulnerability detection tools, and it is widely used in vulnerability detection tasks~\cite{DBLP:journals/toplas/SpotoBEFLMS19, DBLP:journals/tosem/SayarBBT23, DBLP:conf/icse/HoughW0B20}.
We adopt version 1.1 of this benchmark, which contains more data and is suitable for training models.
As a result, the dataset includes 13,041 examples for training and 4,000 examples for validation and testing, respectively.
For code summarization, CodeSearchNet~\cite{DBLP:journals/corr/abs-1909-09436} is a widely used dataset, which includes data from six programming languages. 
We follow existing works~\cite{DBLP:conf/issta/ZengTZLZZ22, DBLP:conf/nips/LuGRHSBCDJTLZSZ21} and use the filtered Java sub-datasets for code summarization, which results in 164,923 examples for training, 5,183 for validation, and 10,955 for testing.

{\color{black}\subsection{Evaluation Metrics} 
We adopt the following metrics for evaluation.

\noindent\textbf{Accuracy.} It is the proportion of correctly predicted instances in the test set, which is used in the task of vulnerability detection.

\noindent\textbf{Precision, Recall, and F1 Score.} 
These three metrics are used for evaluating clone detection.
Precision (P) is the proportion of cloned pairs correctly predicted as cloned to all pairs predicted as cloned.
Recall (R) is the proportion of cloned pairs correctly predicted as cloned to all known real cloned pairs.
F1 is the harmonic mean of precision and recall and it is calculated as: $F_1 = 2 * (P * R) / (P + R)$.

\noindent\textbf{BLEU-4.} 
BLEU is widely used to evaluate the textual similarity between the text generated in generative systems and the ground-truth. BLEU-4~\cite{DBLP:conf/issta/ZengTZLZZ22, DBLP:conf/sigsoft/WangYGP0L22} is a variant of BLEU, where the 4 indicates that four consecutive words (4-gram) are used as the matching unit. 

We fine-tune PTMCs following existing works~\cite{DBLP:conf/nips/LuGRHSBCDJTLZSZ21, DBLP:conf/issta/ZengTZLZZ22}, and Table~\ref{Evaluation} lists the reproduced results.  The results are consistent with the previously reported ones in the original paper, which indicates that the models in our experiments have been adequately fine-tuned. }
\begin{table}[H]
\vspace{-2mm}
    \caption{Evaluation results on pre-trained models of code}
    \label{Evaluation}
    \small
    \vspace{-4mm}
    \renewcommand\arraystretch{0.9}
    \setlength\tabcolsep{7.8pt}
    \centering
    \begin{threeparttable}
    \begin{tabular}{c|c|ccc|c}
    \toprule
    Task & VD     & \multicolumn{3}{c|}{CD} & \multicolumn{1}{c}{CS} \\ \hline
    Metrics    & Acc   & Precision   & Recall  & F1  & BLEU-4        \\ 
    \midrule
    CodeBERT        & 98.70 &96.42  &96.32 &96.32 &18.75   \\ 
    CodeGPT         & 97.45 &96.55  &96.52 &96.52 &15.36   \\
    PLBART          & 99.52 &96.83  &96.83 &96.82 &17.60   \\
    \bottomrule
    \multicolumn{6}{c}{\footnotesize VD: Vulnerability Detection; CD: Clone Detection; CS: Code Summarization} \\
\end{tabular}
	
\end{threeparttable}
\vspace{-4mm}
\end{table}

\subsection{Research Questions}
The goal of this study is to systematically evaluate and compare the performance of the SOTA adversarial attack approaches against various PTMC under different PL tasks, including their effectiveness and efficiency. 
More importantly, we are also curious to know the code qualities of the generated adversarial examples since it is reported that the quality of the generated examples is of significant importance~\cite{DBLP:conf/icse/YangSH022}.
To our best knowledge, it is also the first large-scale investigation towards the quality of the adversarial examples.
Besides, we also investigate whether the context of perturbed identifiers will affect the performance of existing adversarial attack approaches.
We introduce our target RQs in detail as follows:

\textbf{RQ1: (Attacking performance) How do existing adversarial attack approaches perform against different PTMCs under various tasks?} 
{In this RQ, we attempt to thoroughly compare the SOTA adversarial attack approaches based on two criteria~\cite{DBLP:conf/acl/ZengQZZMHZLS21, DBLP:journals/tosem/ZhangFLMZYSLJ22}.}

\noindent\textbf{C1: Effectiveness.} We compare the effectiveness of adversarial attacks according to the \textit{Attack Success Rate} (\textbf{ASR}), which is the percentage of code snippets on which an attack approach can successfully generate adversarial examples, given a code dataset.
A higher ASR indicates a more effective attack.

\noindent\textbf{C2: Efficiency.} We compare the efficiency of adversarial attacks according to two metrics: 
(1) \textit{Average Model Queries} (\textbf{AMQ}). AMQ denotes the number of queries to the attacked model during the generation of adversarial examples, which
is positively related to the attack running time. Too many model queries will be abnormal and suspicious for the attacked party. 
(2) \textit{Average Running Time} (\textbf{ART}). ART is an overall metric of the efficiency of the attack approach. 
It is not only related to the number of model queries, but also to the perturbation strategy. 
For example, the genetic algorithm is more time-consuming than the greedy algorithm~\cite{10.5555/775009.775028}. 

\textbf{RQ2: (Adversarial code quality) What is the quality of adversarial examples generated by adversarial attacks?} 
{Naturalness is crucial in adversarial example generation~\cite{DBLP:conf/icse/YangSH022}, as highlighted by ACCENT~\cite{DBLP:journals/tosem/ZhouZSHCG22}: people will easily argue that if the replaced identifiers are significantly different from the original ones, the summary should be different.
Therefore, they use \textit{cosine similarity} to constrain adversarial examples.}
According to the existing works on the evaluation of perturbation towards text~\cite{DBLP:journals/tkde/WangWWWY23, DBLP:journals/tist/ZhangSAL20} and code~\cite{DBLP:journals/tosem/ZhouZSHCG22, DBLP:conf/icse/YangSH022}, there are two main aspects concerning the quality of the generated examples. First, the number of tokens that are replaced should be as small as possible.
Second, the adversarial tokens need to be as similar as the original ones in terms of their semantics.
In this study, we evaluate the former with \textit{Identifier Change Rate} (\textbf{ICR}) and \textit{Token Change Rate} (\textbf{TCR}), and the latter with \textit{Average Code Similarity} (\textbf{ACS}) and \textit{Average Edit Distance} (\textbf{AED}), following existing studies~\cite{DBLP:journals/tkde/WangWWWY23, DBLP:journals/tist/ZhangSAL20,DBLP:journals/tosem/ZhouZSHCG22, DBLP:conf/icse/YangSH022}.
The calculation of these four metrics are as follows.
(1) For $k$ adversarial examples, if there are in total $m_i$ identifiers in the $i_{th}$ code snippet and $n_i$ identifiers have been changed in the adversarial examples, then ICR is evaluated as ${\textstyle \sum_{1}^{k}n_i}/{\textstyle \sum_{1}^{k}m_i}$;
(2) Beyond the identifiers, the source code may contain other code tokens such as keywords, operators, etc.
TCR is the ratio of the changed tokens in the adversarial example to the total number of tokens in the entire code.
(3) We use the cosine similarity to reflect the code similarity before and after the perturbations are performed. In particular, ACS is computed based on the embeddings that vectorized from the source code by CodeBERT;
(4) AED reflects the character-level token differences, which is the number of times a token needs to be edited at the character level in order to transform into another. 
In general, a high-quality adversarial example should preserve lower ICR, AED, and TCR while the ACS should be higher. 

\textbf{RQ3: (Context of perturbed identifiers) How do the contexts of the perturbed identifiers affect the adversarial attacking performance?} 
Existing adversarial perturbations tend to treat all identifiers equally, which leads to a large search space and might also compromise the attacking efficiency. 
To reduce the search overhead, we aim to explore the impact of the contexts of different identifiers on the attacking results in this RQ.
In particular, we regard the statements where the identifiers reside as contexts and investigate whether perturbing identifiers residing at different contexts will affect the attacking effectiveness. 
In this study, we select the top five statements that are commonly used in code~\cite{DBLP:journals/jss/QiuLBS17} for investigation, which are {\mycode Return}, {\mycode If}, {\mycode Throw}, {\mycode Try}, and {\mycode For} statements.
In addition to these types of statements, we also investigate the impact of merely modifying method names and the parameters to verify whether the models are vulnerable to such changes. 
\textit{\textbf{We refer to them as {\mycode Method} in this study}}.
We use ASR to observe the impact. 
Particularly, we choose two attacks with the highest ASR, which are MHM and WIR-Random. 
Finally, we use CodeBERT as the target model because it is the most studied PTMC to date.

\vspace{-2mm}
\subsection{Settings of Attacks}
We use the trained models as introduced in Section~\ref{sec:model} as the attack targets and adapt the original code of the five attack approaches in this study. In particular, we only make limited modifications on the code, specifically focusing on the data loading and a few parameters (\eg~the candidate identifiers as mentioned in Section~\ref{sec:attack_method}), to serve for the need of processing our selected datasets.
We use all the test set as the target instances (\ie~in total 4,000) for attacks on vulnerability detection and clone detection.
For the code summarization task, we randomly select 4,000 examples from the test set as instances used for attacks to align with the number of the target instances used in the other two tasks.
Meanwhile, it is beneficial for our study to explore the differences between the robustness of models for different tasks under the same scale of adversarial attacks.
When evaluating the attack approaches based on identifier substitution, we skip source programs without identifiers. 
Besides, we also skip the instances that are classified incorrectly by the model to mitigate the effect of model performance. 
Such settings are commonly used in adversarial attacks~\cite{DBLP:conf/icse/YangSH022, DBLP:conf/aaai/ZhangLLMLJ20}. 
Although we exclude a small proportion of instances, our study is large-scale. 
In particular, we perform attacks on more than 150,000 target programs with over 100 million queries to various PTMC models.

	\vspace{-2mm}
\section{Empirical Results}
In this section, we present the results of our empirical studies. 
\vspace{-2mm}
\subsection{Attack Performance (RQ1)}
\subsubsection{Effectiveness}
We perform experiments on the five attack approaches and measure their \textit{Attack Success Rate} (ASR), and the results are shown in Table~\ref{table:ASR}. 
Generally, all the three target models can be easily attacked under the three different tasks. 
In particular, MHM can achieve the highest ASR (\ie~57.83\%) averaged over all the experiments, followed by WIR-Random (\ie~38.77\%). 
Based on the results, we make the following observations. 

First, \textit{random substitution is more effective than the other perturbation strategies}.
Specifically, both MHM and WIR-Random adopt the strategy of random substitution while ALERT perturbs identifiers based on context-aware prediction.
Consequently, MHM and WIR-Random outperform ALERT by 184.60\% and 90.80\%, respectively.
Meanwhile, such outperformance can be observed for all the three tasks, which reflects that random substitution is the most effective strategy to mislead pre-trained models.
On the contrary, StyleTransfer is less effective.
We conjecture the behind reason is that existing trained clone detection models are more robust to various code transformation strategies.
For example, the cloning method summarized by
Walker~\emph{et~al.}~\cite{10.1145/3381307.3381310} includes adding/deleting code snippets and reordering statements, which is very similar to the strategies as adopted by StyleTransfer. 
Therefore, the clone detection model can learn sufficient code transformation features on such code clone pairs, thus being robust to StyleTransfer.  

Second, \textit{the models are more robust against  adversarial attacks under the understanding tasks than the generation tasks}.
In particular, we observe that 
the ASR of the five attacks on the code summarization model is higher than that of the clone detection and vulnerability detection. 
The average ASRs of clone detection and vulnerability detection are 24.39\% and 22.62\%, much lower than that of code summarization, which is 52.25\%.
Among them, MHM achieves an ASR over 90\% on the three code summarization models on average, which shows that these models can be easily attacked under the task of code summarization, and output completely irrelevant summaries compared to their original outputs.

Third, \textit{among the different pre-trained models, CodeGPT is more resistant to various attacks.}
CodeGPT achieves the lowest ASR in 11/15 of the experiments (five attacks for three tasks).
The average ASR over the three tasks of the five attacks on CodeGPT is 25.51\% as shown in the last column of Table~\ref{table:ASR}, which is lower than that of CodeBERT by 37.88\% and PLBART by 35.87\% respectively.  

\begin{table}[]
    \small
    \caption{Attack success rate on pre-trained models of code}
    \vspace{-4mm}
    \label{table:ASR}
    \renewcommand\arraystretch{0.9}
    \setlength\tabcolsep{2.0pt}
    \begin{tabular}{cr|rcccc|c}
    \toprule
    \multicolumn{2}{c}{\multirow{1}{*}{Attack Approach}} & \multirow{1}{*}{MHM} &\multirow{1}{*}{ACCENT} & \multirow{1}{*}{ALERT} & \multirow{1}{*}{WIR} &\multirow{1}{*}{StyleTransfer} &\multirow{1}{*}{Avg}\\
    \midrule
    
    \multirow{3}{*}{CD} 
    & CodeBERT  & 47.13 & 21.58  & 14.48 & 35.00 & 0.42 &23.72 \\
    & CodeGPT   & 43.90 & 23.55  & 6.59  & 35.43 & 0.27 &21.95 \\
    & PLBART    & 45.89 & 51.64  & 9.10  & 30.23 & 0.68  &27.51\\
    \midrule
    \multirow{3}{*}{VD} 
    & CodeBERT  & 57.12  & 31.10 & 4.23 & 18.17 & 21.97 &26.52 \\
    & CodeGPT   & 29.68  & 24.95 & 4.75 & 13.11 & 9.04 &16.31 \\
    & PLBART    & 25.09  & 35.07 & 17.28 & 22.23 & 25.58 &25.05 \\
    \midrule
    \multirow{3}{*}{CS}
    & CodeBERT  & 93.80  & 64.85 & 48.58 & 79.84 & 29.97  &63.41\\
    & CodeGPT   & 87.11  & 21.17 & 23.50 & 52.90 & 6.71  &38.28\\
    & PLBART    & 90.76  & 53.73 & 54.37 & 62.00 & 14.45  &55.06\\   
    \midrule
    \multicolumn{2}{c}{Average Number} &57.83 	&36.40 	&20.32 	&38.77 	&12.12	&
                                 \\
    \bottomrule
    \multicolumn{8}{l}{CD:Clone Detection; VD:Vulnerability Detection; CS:Code Summarization}
    \end{tabular}
    \vspace{-4mm}
\end{table}
\begin{center}
\vspace{-2mm}
\begin{tcolorbox}[colback=gray!10, colframe=black, width=8.5cm, arc=1mm, boxrule=0.5pt, left=1mm, right=1mm, top=0.5mm,bottom=0.5mm]
\textbf{Finding 1: } 
Pre-trained models with excellent performance can be easily misled by various adversarial attacks. 
In particular, random strategies are more effective; models for the generation tasks are less robust compared to understanding tasks; and CodeGPT is in general more resistant to various attacks. 
\end{tcolorbox}
\end{center}

\subsubsection{Efficiency}
\renewcommand\theadfont{\small}
\begin{table*}[]
    \small
    \caption{\textit{Average Model Queries} (AMQ) and \textit{Average Running Time} (ART) on attacking CodeBERT, CodeGPT, and PLBART}
    \label{table:AMQ&ART}
    \renewcommand\arraystretch{0.9}
    \setlength\tabcolsep{3.0pt}
    \vspace{-4mm}
    \begin{tabular}{cc|rrrrr|rrrrr}
    \toprule
    \multicolumn{2}{c|}{\multirow{2}{*}{Attack Approach}} &  \multicolumn{5}{c|}{Average Model Queries (AMQ)}  & \multicolumn{5}{c}{Average Running Time (ART) (min)} \\ \cline{3-12}
    & & \multirow{1}{*}{MHM} &\multirow{1}{*}{ACCENT} & \multirow{1}{*}{ALERT} & \multirow{1}{*}{WIR-Random} & \multirow{1}{*}{StyleTransfer} & \multirow{1}{*}{MHM} &\multirow{1}{*}{ACCENT} & \multirow{1}{*}{ALERT} & \multirow{1}{*}{WIR-Random} &\multirow{1}{*}{StyleTransfer} \\ 
    \midrule
    \multirow{3}{*}{\thead{Clone \\ Detection}} 
    & CodeBERT 	&1,884.43 	&196.09	&\textbf{2,263.53} &247.65	&498.12	 	&5.69    &2.95 	&4.49	&0.63 	&\textbf{9.25} 	   \\
    & CodeGPT  	&2,040.39	&206.35 	&\textbf{2,596.18} &250.58 &498.83 	 	&6.94   &3.13   &5.89   &0.27   &\textbf{14.09}      \\
    & PLBART 	&2,040.01	&157.03 	&\textbf{2,549.49} &256.27 &496.97 		&7.79 	&2.36 	&8.39	&0.33 	&\textbf{18.35} 	  \\
    \midrule
    \multirow{3}{*}{\thead{Vulnerability \\ Detection}} 
    & CodeBERT  &1,877.43	&235.83 	&\textbf{2,718.57}	&276.30	&397.33 		& 1.75	&5.09 	&2.02	&0.24 	&\textbf{20.00} 	  \\
    & CodeGPT   &2,497.18 	&228.29 	&\textbf{2,798.29} &282.98 &459.41 		&2.72	&5.33 	&1.55	&0.23 	&\textbf{21.42} 	  \\
    & PLBART    &2,446.50	&256.31	&\textbf{2,592.32}	&277.02 &382.72 		& 3.37	&4.66	&3.65	&0.27 	&\textbf{20.07} 	 \\
    \midrule
    \multirow{3}{*}{\thead{Code \\ Summarization} }
    & CodeBERT  &395.01	&36.49 	&\textbf{565.63} 	&90.31 &368.71 		&1.38 	&0.21 &	2.30 	&0.61 	&\textbf{7.19} 	  \\
    & CodeGPT   &756.67	&76.15 	&\textbf{938.39} 	&126.64 	&469.58 	  &8.29 	&0.80 	&8.78 	&1.83 	&\textbf{11.80} 	 \\
    & PLBART    &\textbf{587.51}	&46.27 	&488.82 	&109.07 &438.67 		& 1.62 	&0.26 	&1.70 	&0.30 &	\textbf{10.34} 	  \\
    \midrule
    \multicolumn{2}{c|}{Average Number}	&1,613.90	&159.87 	&\textbf{1,945.69} 	&212.98 	&445.59 	 &4.39 	&2.75 	&4.31 	&0.52 	&\textbf{14.72} 		\\                                     
    \bottomrule
    \end{tabular}
    \vspace{-2mm}
\end{table*}

Table~\ref{table:AMQ&ART} shows the results with respect to AMQ and ART.
Via analyzing these two metrics in conjunction with ASR, 
we make the following observations.

First, \textit{the efficiency of different attacks varies greatly.} 
The average AMQ of ALERT and MHM is 1,945.69 and 1,613.90 respectively, while that of WIR-Random and ACCENT is only 212.98 and 159.87. 
Such differences are caused by the characteristics of the attack approaches themselves. In particular, MHM employs a large number of iterations while StyleTransfer only transfers the target code for a limited number of times to maintain the naturalness of the code.
Besides, ALERT uses the genetic algorithm with multiple iterations, which tends to repeatedly replace the same identifier, while WIR-Random and ACCENT only replace identifiers sequentially according to their importance calculated by the algorithm, and they will not repeat replacing identifiers.
For the same attack, the efficiency varies on different tasks as well. 
Specifically, the average AMQ of the five attack approaches on the three PTMCs is 1,078.80 and 1,181.77 on  clone detection and  vulnerability detection, but this value is 366.26 on code summarization. 
Further analysis reveals that it is caused by the low robustness of  code summarization models (Finding~1). 
The high ASR of the attack approaches on code summarization models means that attacks can terminate early without performing all iterations or visiting all replaceable variables.

\begin{figure}[t!]
    \centering

\begin{subfigure}{0.15\textwidth}  
  \centering
    \includegraphics[width=1\linewidth]{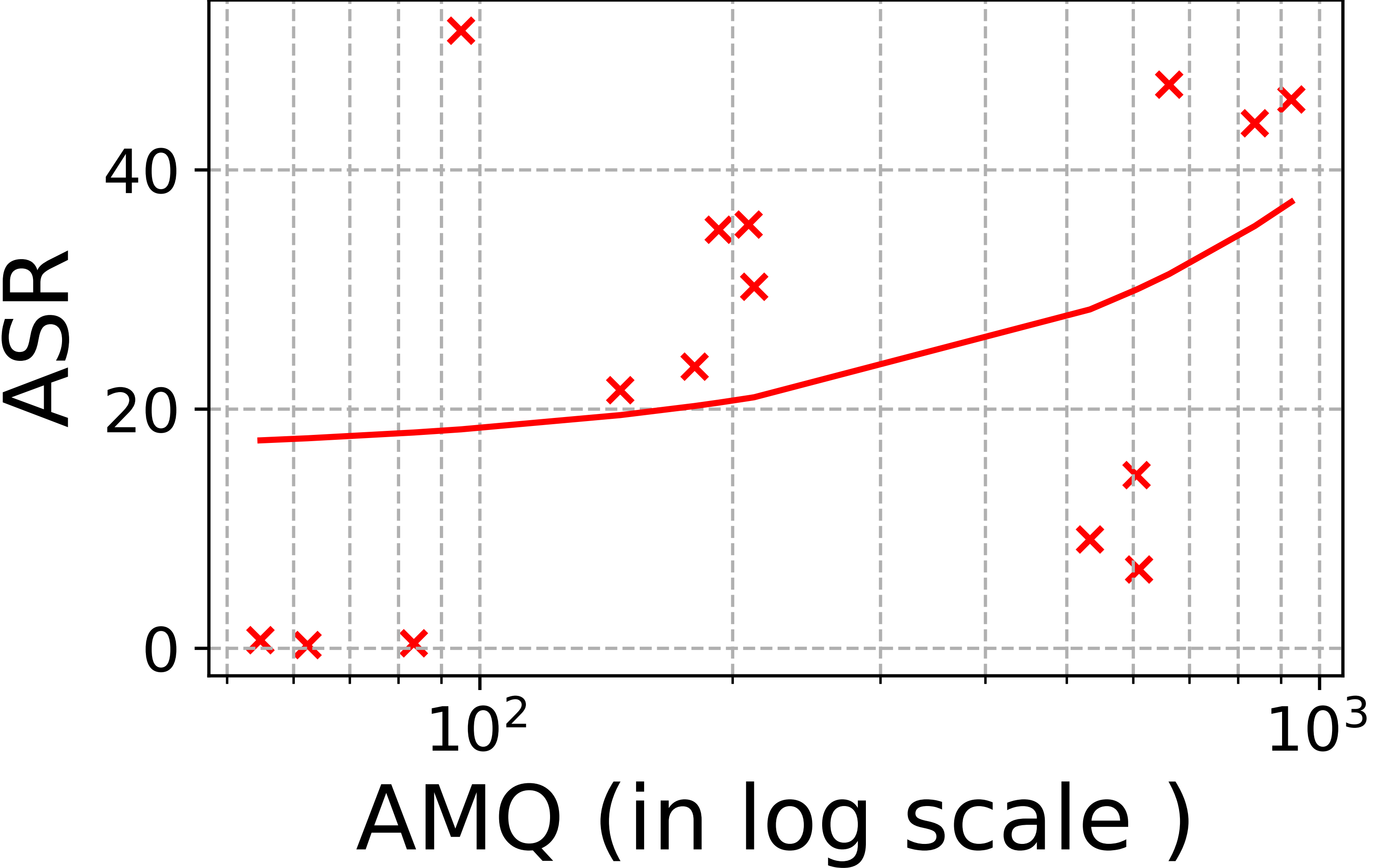}
    \caption{CD} \label{fig:scatter_ASR_AMQ_clone}
\end{subfigure}
\begin{subfigure}{0.15\textwidth}  
  \centering
    \includegraphics[width=1\linewidth]{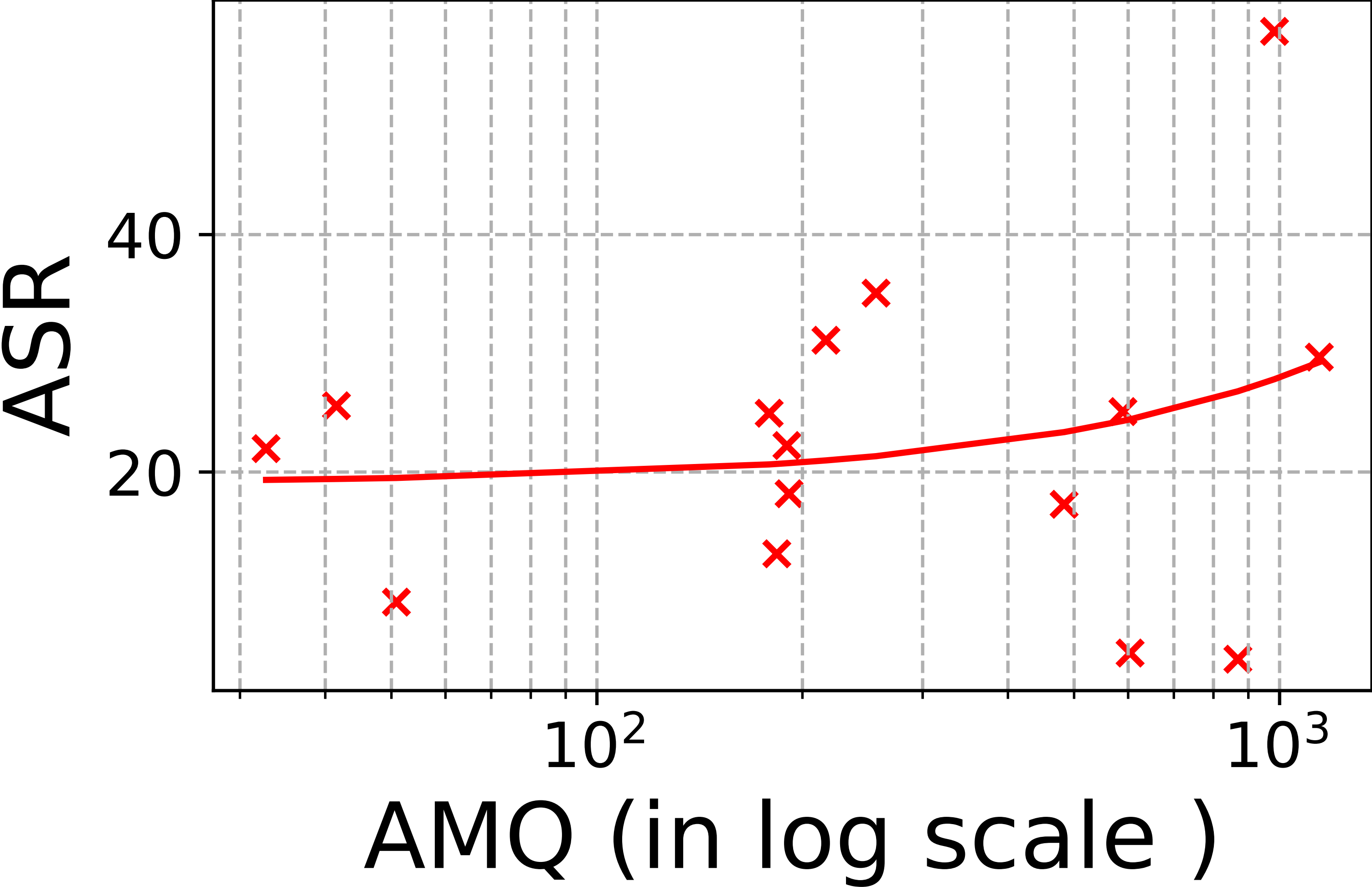}
    \caption{VD} \label{fig:scatter_ASR_AMQ_vul}
\end{subfigure}
\begin{subfigure}{0.15\textwidth}  
  \centering
    \includegraphics[width=1\linewidth]{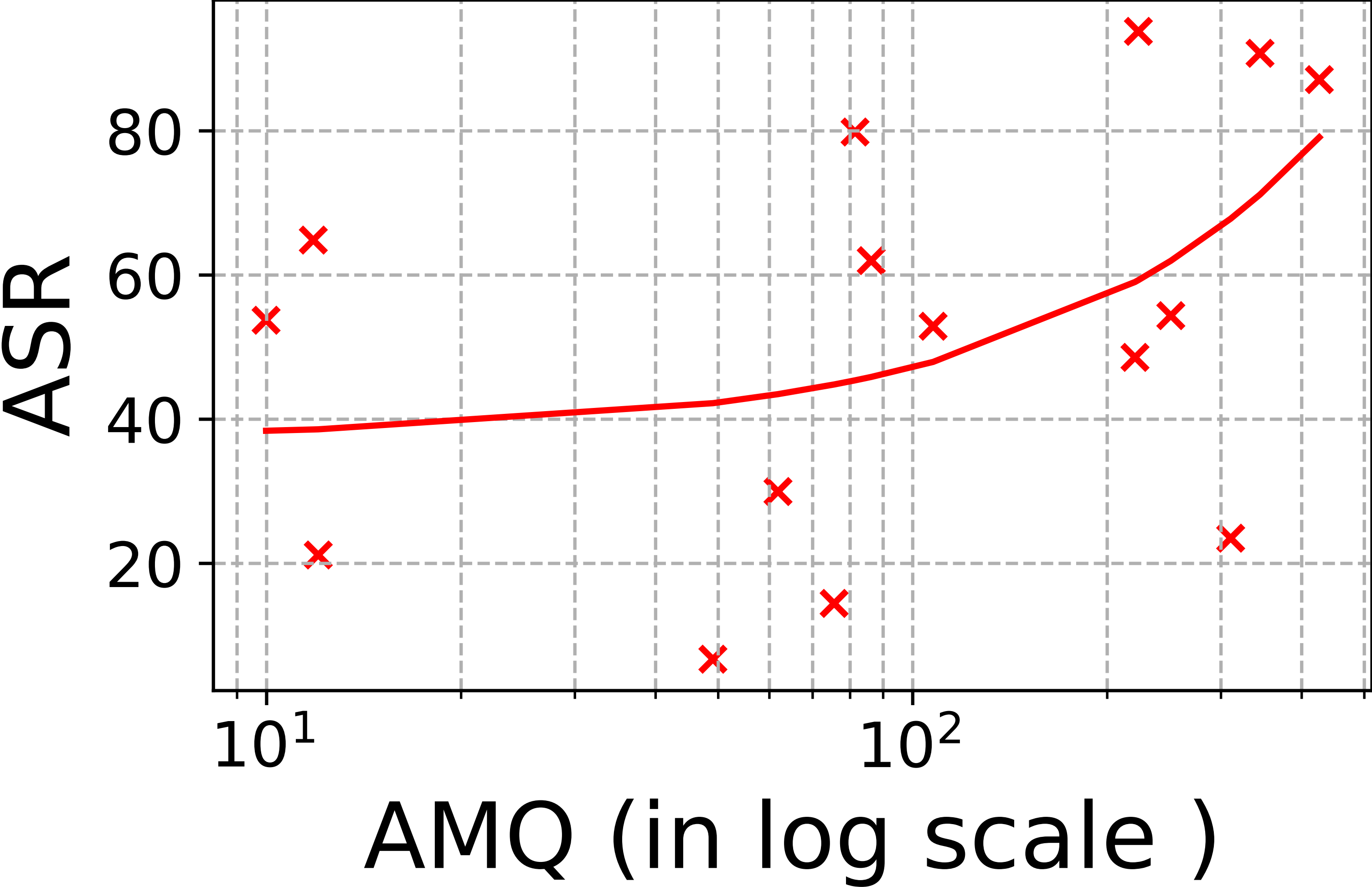}
    \caption{CS} \label{fig:scatter_ASR_AMQ_sum}
\end{subfigure}
\begin{subfigure}{0.15\textwidth}  
  \centering
    \includegraphics[width=1\linewidth]{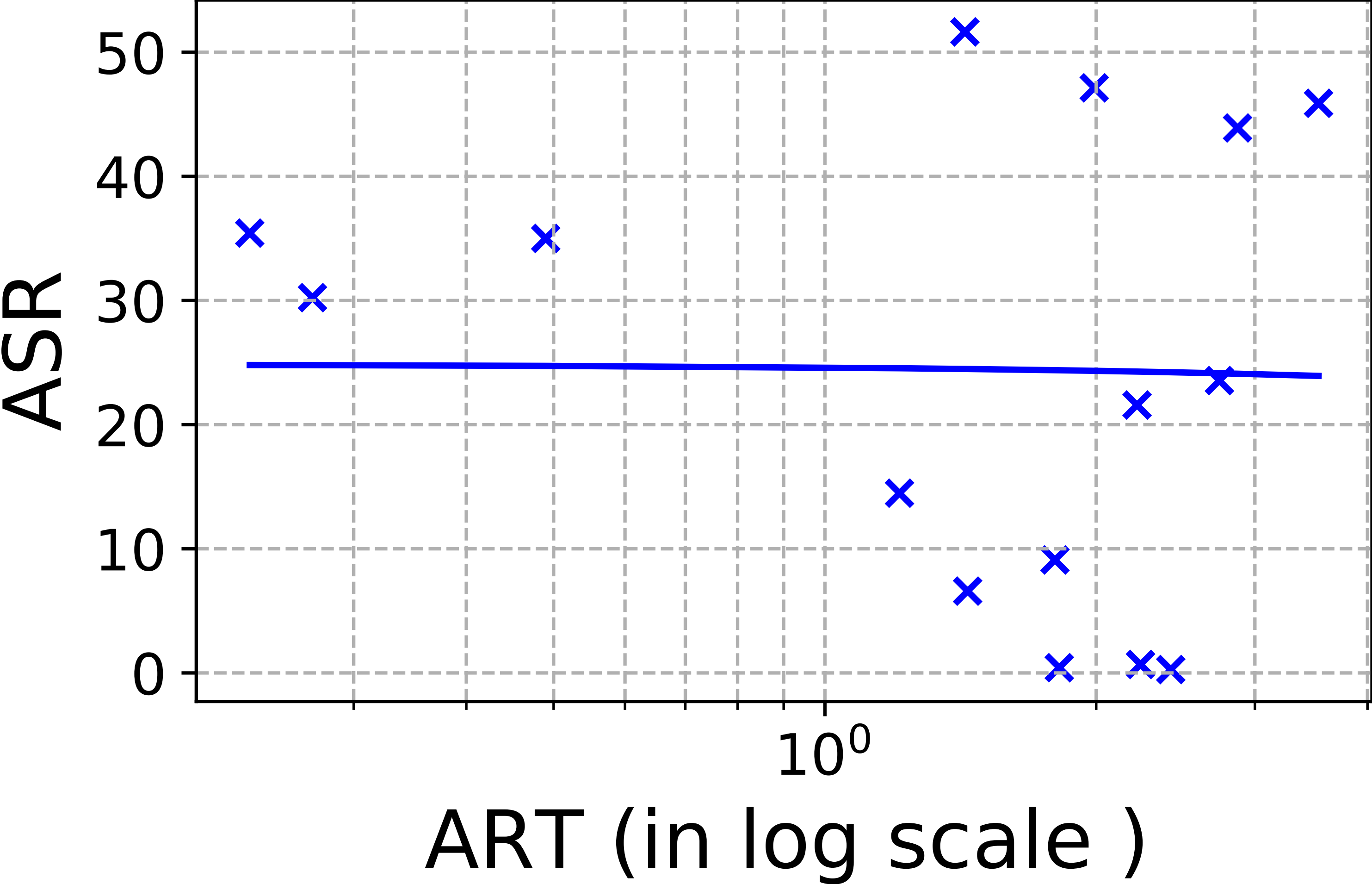}
    \caption{CD} \label{fig:scatter_ASR_ART_clone}
\end{subfigure}
\begin{subfigure}{0.15\textwidth}  
  \centering
    \includegraphics[width=1\linewidth]{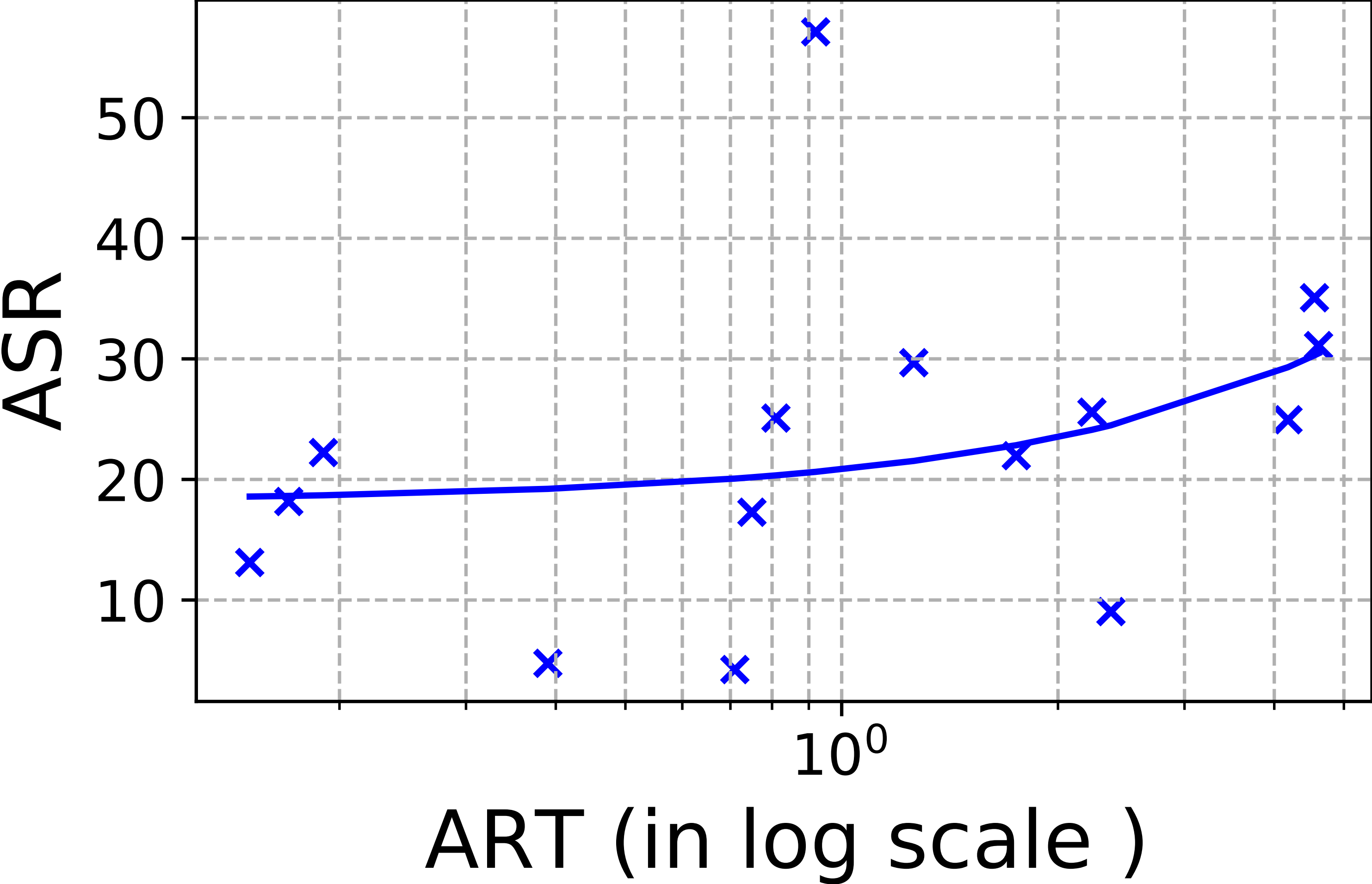}
    \caption{VD} \label{fig:scatter_ASR_ART_vul}
\end{subfigure}
\begin{subfigure}{0.15\textwidth}  
  \centering
    \includegraphics[width=1\linewidth]{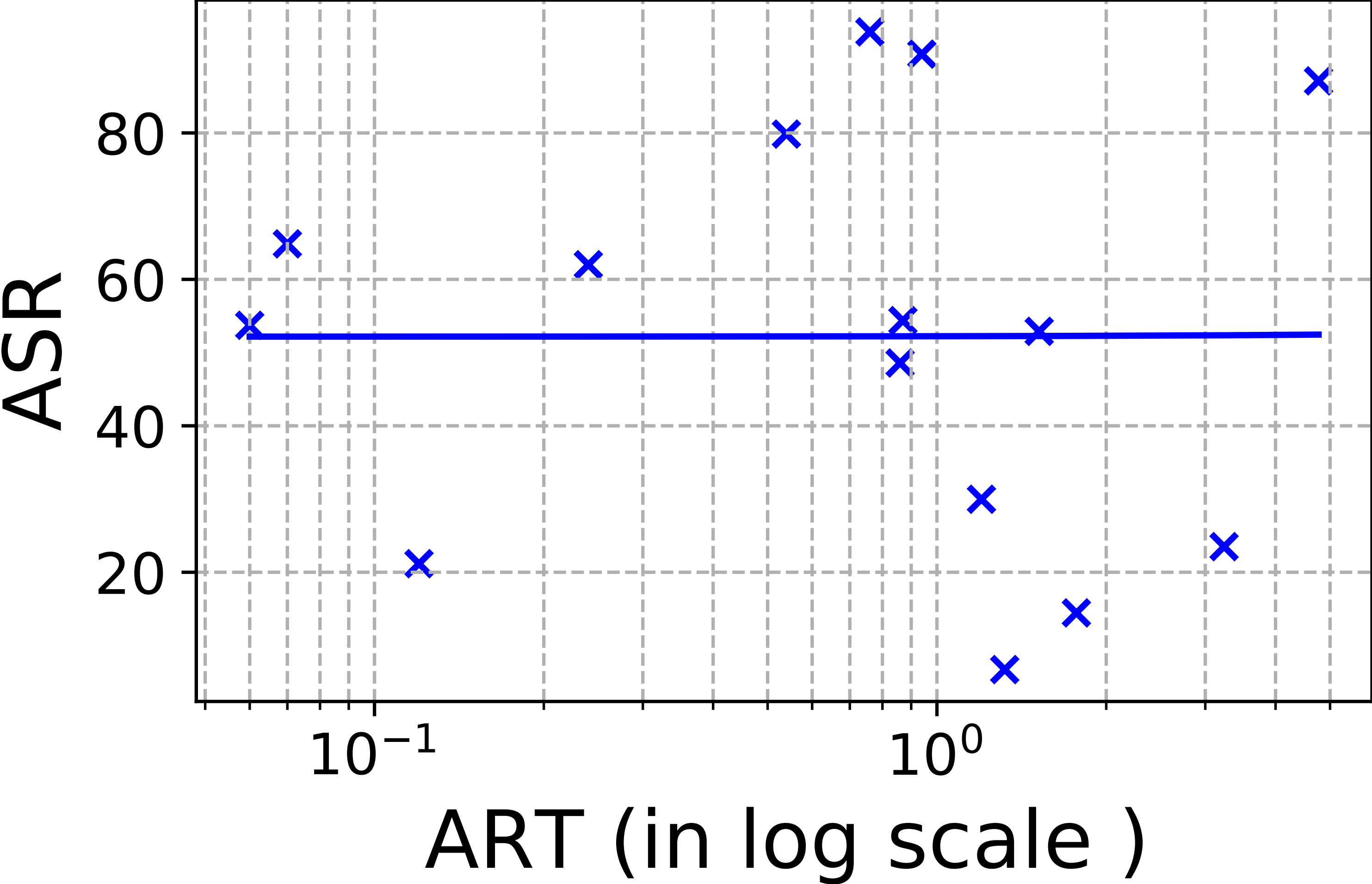}
    \caption{CS} \label{fig:scatter_ASR_ART_sum}
\end{subfigure}
\vspace{-4mm}
\caption{
The correlation between AMQ\&ART and ASR
\label{fig:AMQ&ART}
}
\vspace{-4mm}
 \end{figure}
Second, \textit{the number of model queries is positively correlated with the attack successful rate in general.}
Figure~\ref{fig:AMQ&ART} depicts the correlation between AMQ and ASR. 
As it reveals, attacks with a higher ASR often require a larger number of AMQ. 
For example, the MHM with the highest ASR has an average AMQ of 1,613.90 across all models, while the corresponding values of ACCENT and StyleTransfer with a lower ASR are 159.87 and 445.59 respectively. 
However, the running time (ART) is not necessarily positively correlated with ASR.
Specifically, although StyleTransfer queries the models for less times than MHM (445.59 vs 1,613.90 on average), it takes much longer time for StyleTransfer to process the queries than MHM (14.72 mins vs 4.39). 
As a result, the metric ART is not positively correlated with ASR as shown in Figure~\ref{fig:AMQ&ART}. 
It is because an attack approach often contains additional time consumption besides querying the model. For instance, StyleTransfer usually spends a lot of time on code transformation.

\begin{figure}[htbp]
    \centering

\begin{subfigure}{0.15\textwidth}  
  \centering
    \includegraphics[width=1\linewidth]{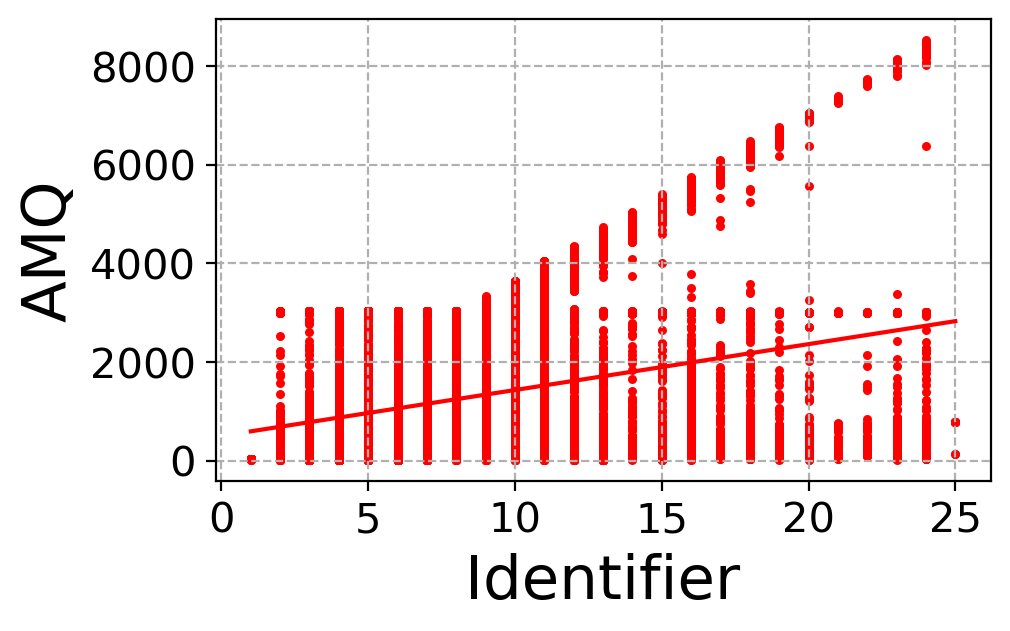}
    \caption{CD} \label{fig:scatter_AMQ_Identifier_clone_no_log}
\end{subfigure}
\begin{subfigure}{0.15\textwidth}  
  \centering
    \includegraphics[width=1\linewidth]{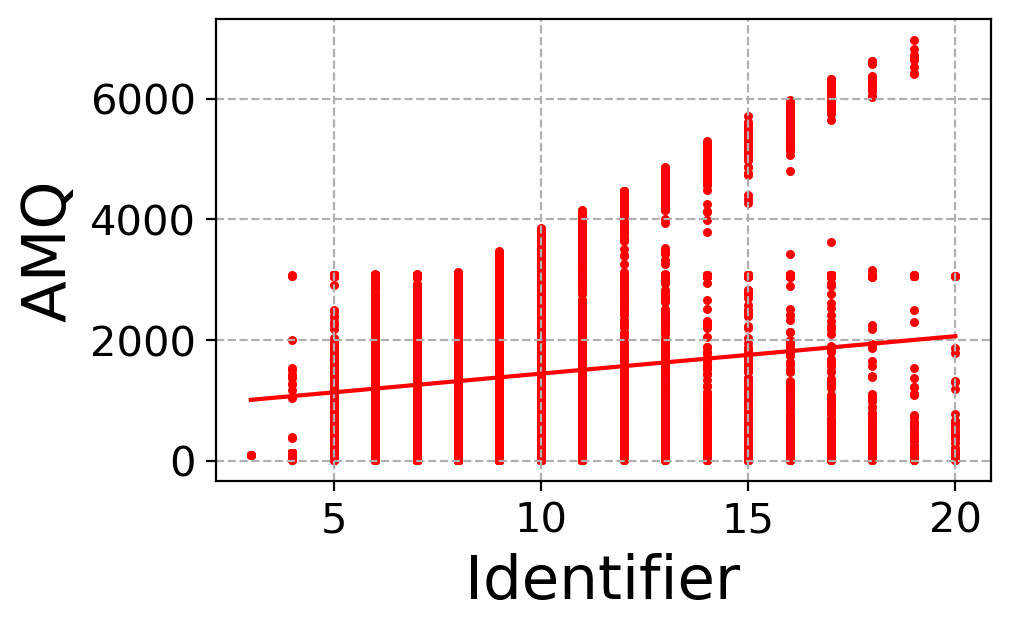}
    \caption{VD} \label{fig:scatter_AMQ_Identifier_vul_no_log}
\end{subfigure}
\begin{subfigure}{0.15\textwidth}  
  \centering
    \includegraphics[width=1\linewidth]{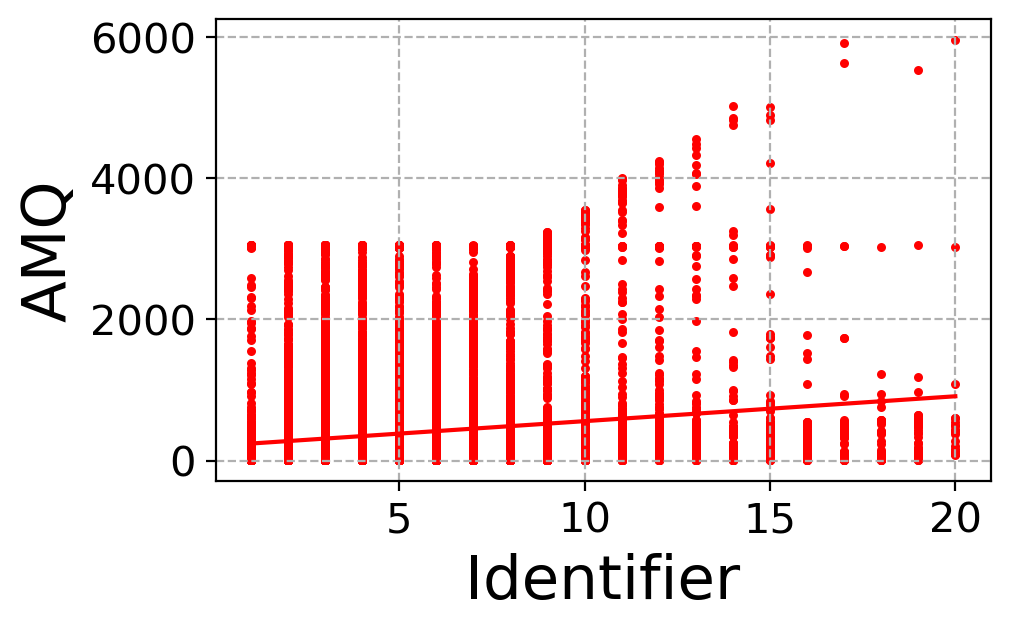}
    \caption{CS} \label{fig:scatter_AMQ_Identifier_sum_no_log}
\end{subfigure}
\begin{subfigure}{0.15\textwidth}  
  \centering
    \includegraphics[width=1\linewidth]{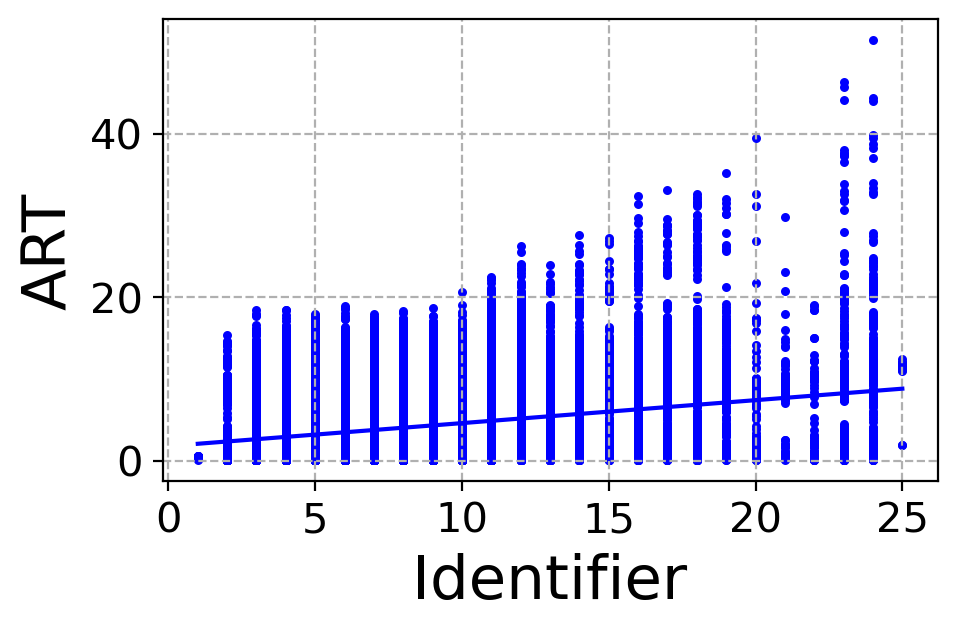}
    \caption{CD} \label{fig:scatter_ART_Identifier_clone_no_log}
\end{subfigure}
\begin{subfigure}{0.15\textwidth}  
  \centering
    \includegraphics[width=1\linewidth]{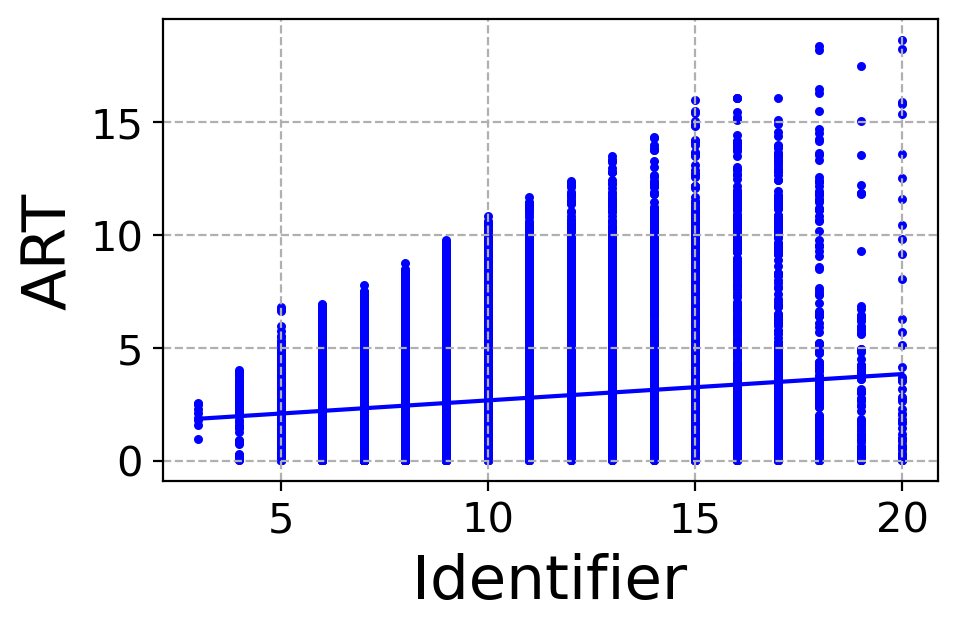}
    \caption{VD} \label{fig:scatter_ART_Identifier_vul_no_log}
\end{subfigure}
\begin{subfigure}{0.15\textwidth}  
  \centering
    \includegraphics[width=1\linewidth]{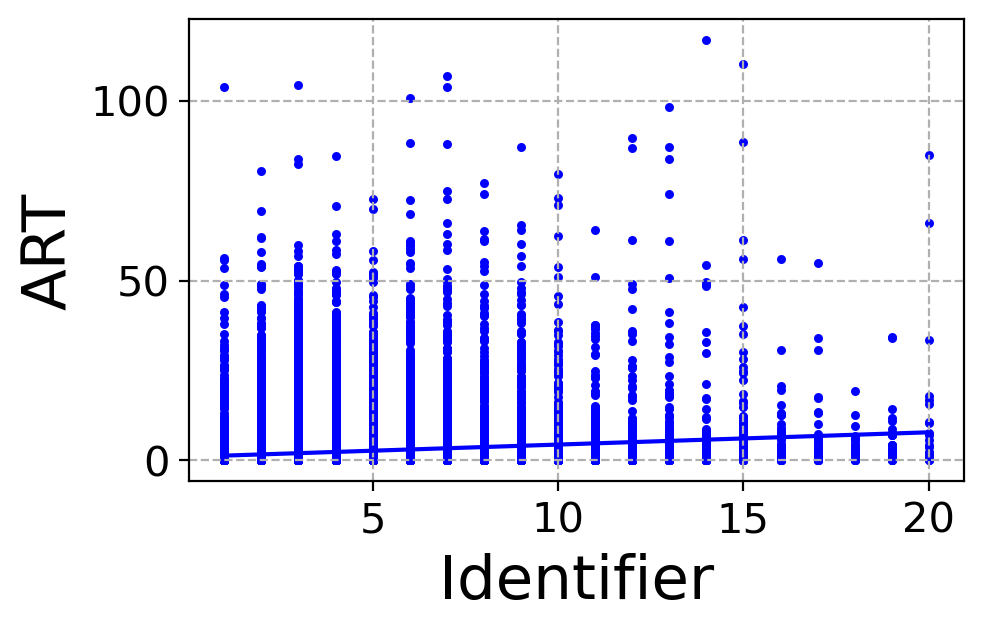}
    \caption{CS} \label{fig:scatter_ART_Identifier_sum_no_log}
\end{subfigure}
\vspace{-4mm}
\caption{
The correlation between AMQ\&ART and the number of identifiers in the target program
\label{fig:AMQ&Identifier}
}
\label{fig:scatter2}
\vspace{-6mm}
 \end{figure}
\begin{figure}[htbp]
    \centering

\begin{subfigure}{0.15\textwidth}  
  \centering
    \includegraphics[width=1\linewidth]{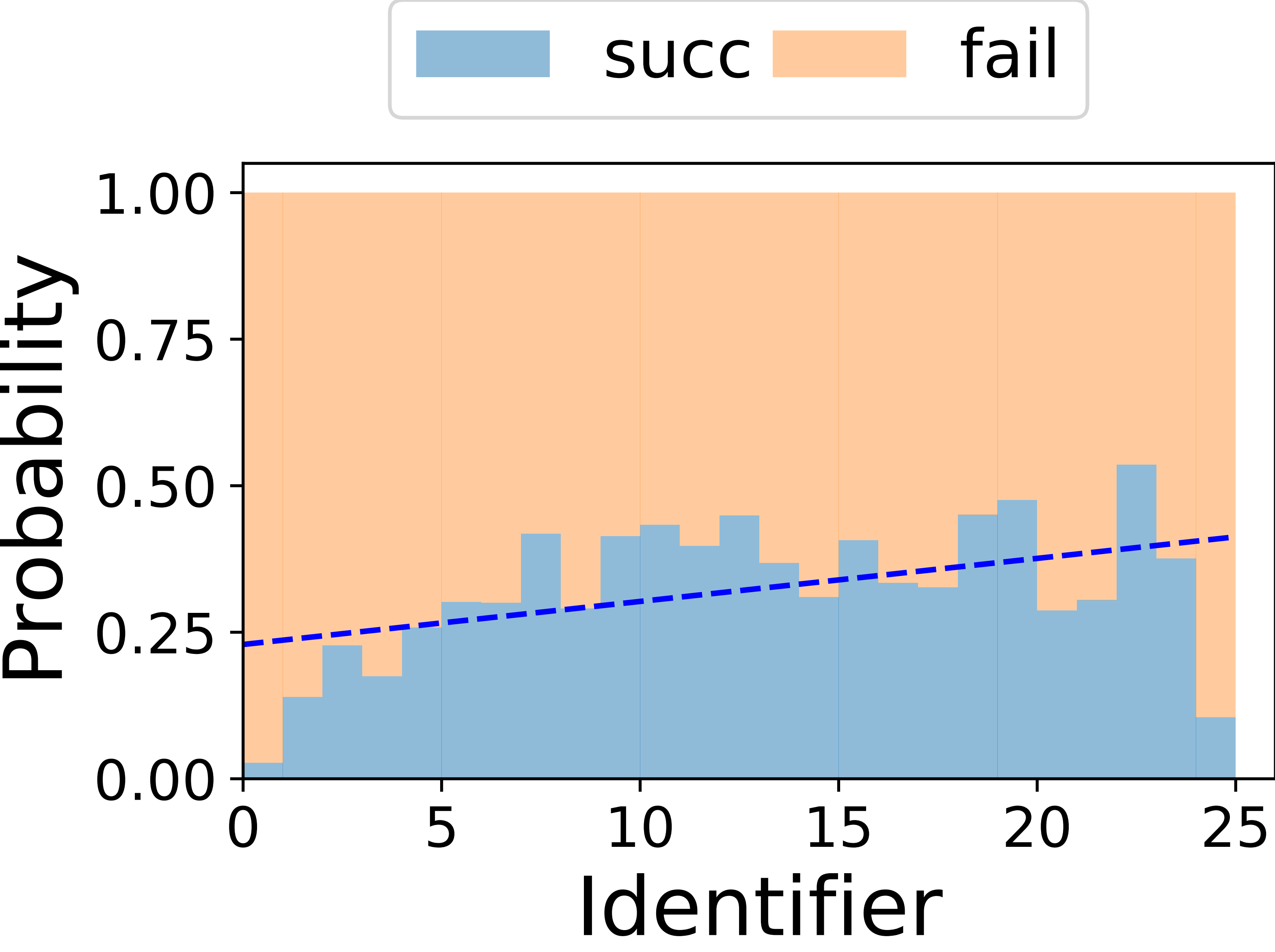}
    \caption{CD} \label{fig:bar_ASR_Identifier_clone}
\end{subfigure}
\begin{subfigure}{0.15\textwidth}  
  \centering
    \includegraphics[width=1\linewidth]{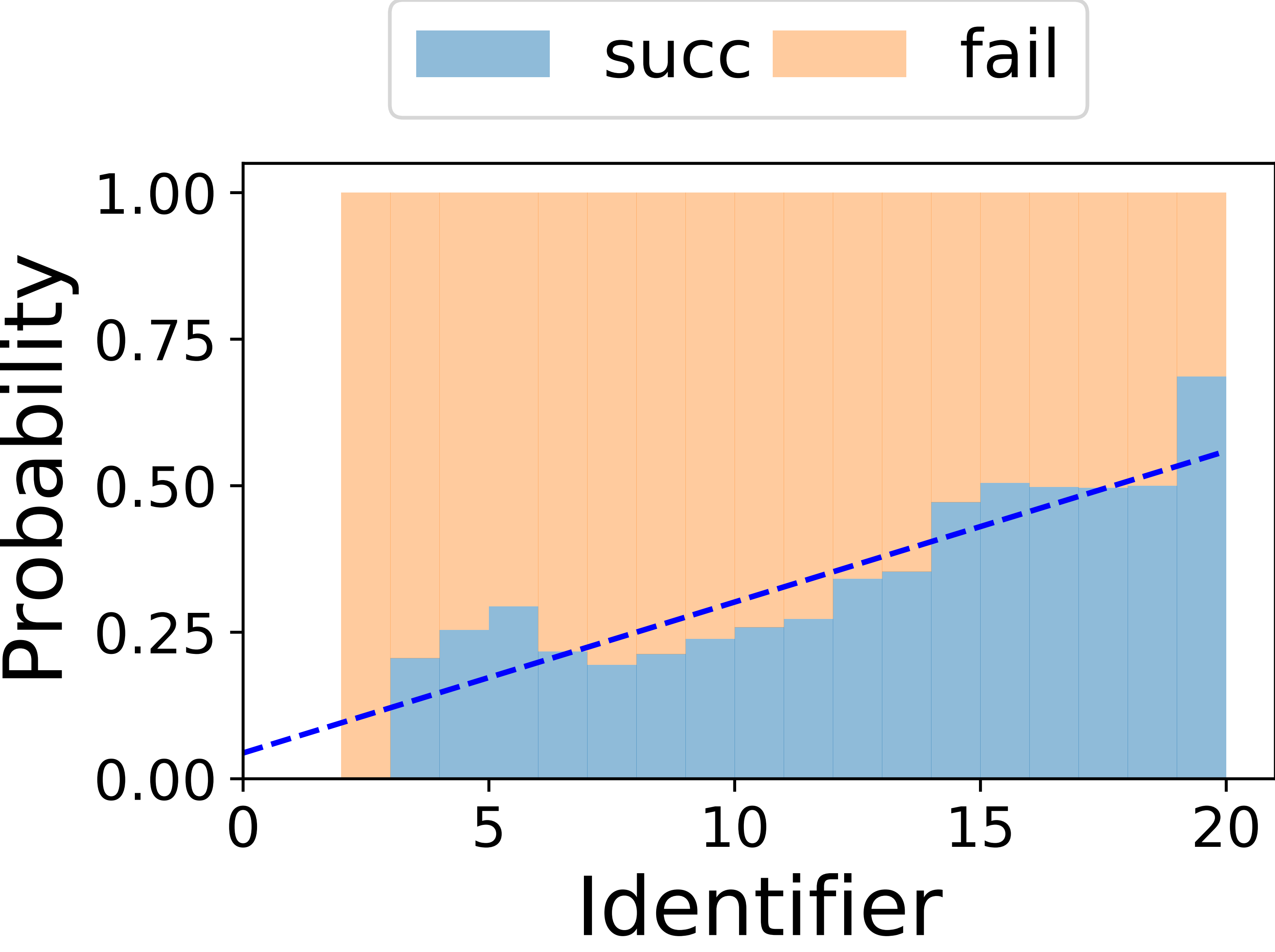}
    \caption{VD} \label{fig:bar_ASR_Identifier_vul}
\end{subfigure}
\begin{subfigure}{0.15\textwidth}  
  \centering
    \includegraphics[width=1\linewidth]{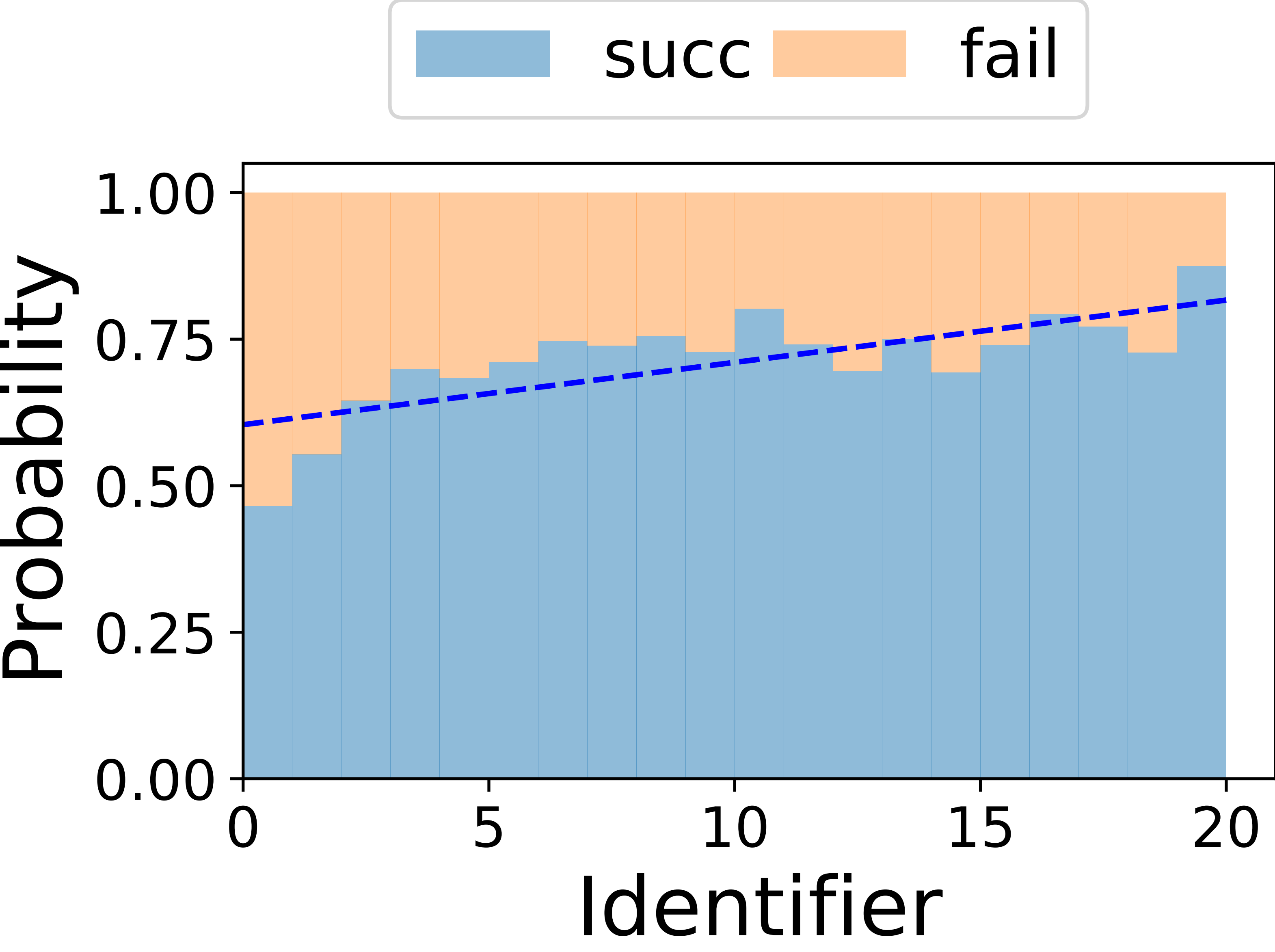}
    \caption{CS} \label{fig:bar_ASR_Identifier_sum}
\end{subfigure}
\vspace{-4mm}
\caption{The correlation between ASR and the number of identifiers in the target program
\label{fig:ASR&Identifier}
}
\label{fig:histogram}
\vspace{-4mm}
 \end{figure}
Third, \textit{the efficiency of various attacks is affected by the total number of identifiers that can be extracted from the program.}
As shown in Figure~\ref{fig:AMQ&Identifier}, AMQ and ART always increase with the number of identifiers, and this trend holds for all the three tasks. 
Such a trend arises from the fact that the number of replaceable identifiers plays the fundamental role in attacks. 
Specifically, both ALERT and WIR utilize the greedy algorithm to iterate over all identifiers. 
Therefore, in the worst case, where the attack fails, its theoretical number of queries is the product of the number of identifiers and the number of potential candidates. 
We further explore the correlation between the number of identifiers in the target program and the attack successful rate, and the results in Figure~\ref{fig:ASR&Identifier} show that more identifiers can in general lead to higher attack successful rates. 
\begin{center}
\begin{tcolorbox}[colback=gray!10, colframe=black, width=8.5cm, arc=1mm, boxrule=0.5pt, 
				  left=1mm, right=1mm, top=0.5mm,bottom=0.5mm]
\textbf{Finding 2: } 
There is a trade-off between the effectiveness and efficiency for adversarial attacks. 
Attacking with higher successful rates often requires a larger number of model queries. 
Besides, the efficiency of attack is also affected by the number of identifiers in the target program.
\end{tcolorbox}
\end{center}

\subsection{Adversarial Code Quality (RQ2)}
\begin{figure}[htbp]
    \centering

\begin{subfigure}{0.15\textwidth}  
  \centering
    \includegraphics[width=1\linewidth]{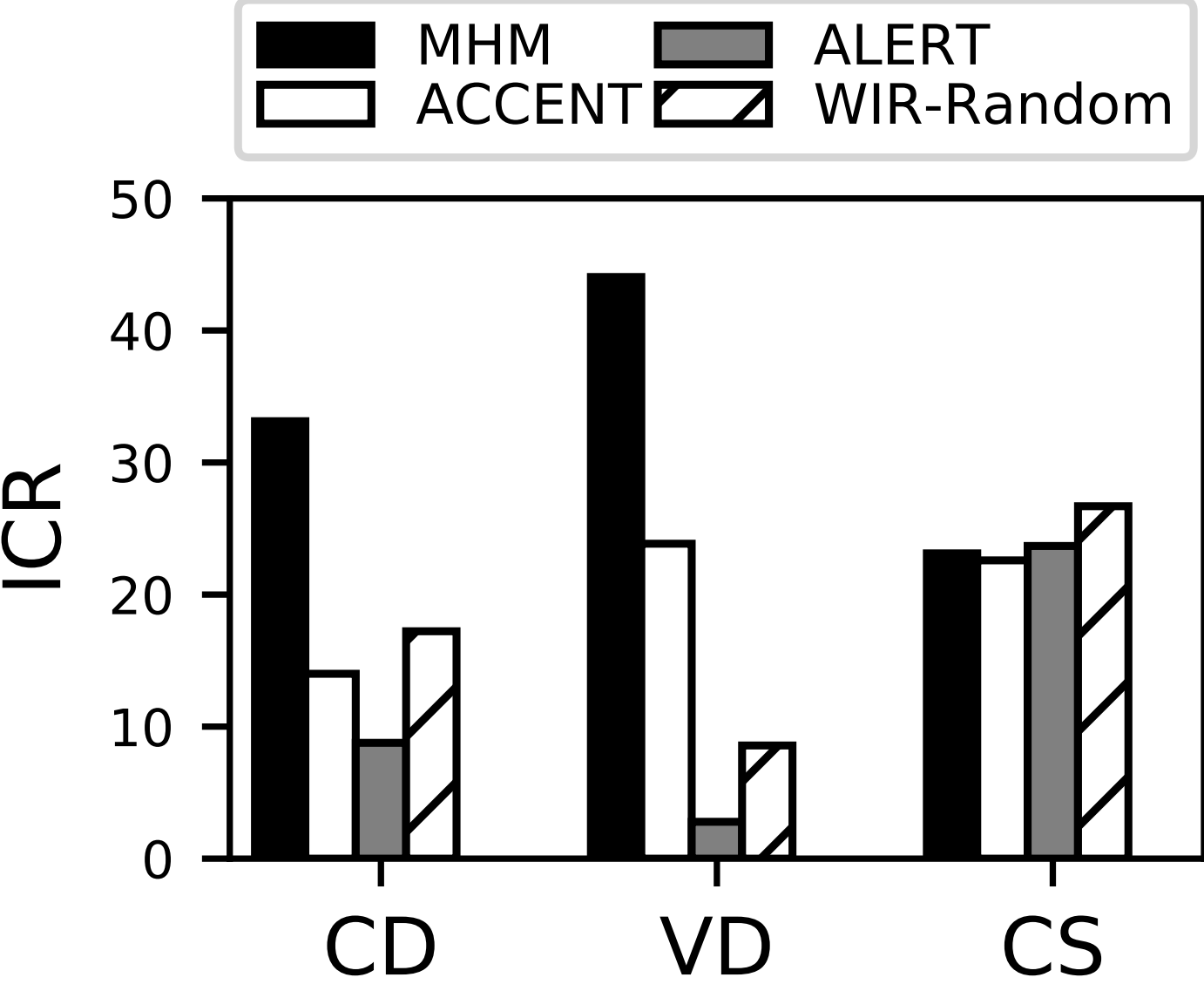}
    \caption{ICR-CodeBERT} \label{fig:ICR_CodeBERT}
\end{subfigure}
\begin{subfigure}{0.15\textwidth} 
  \centering
  \includegraphics[width=1\linewidth]{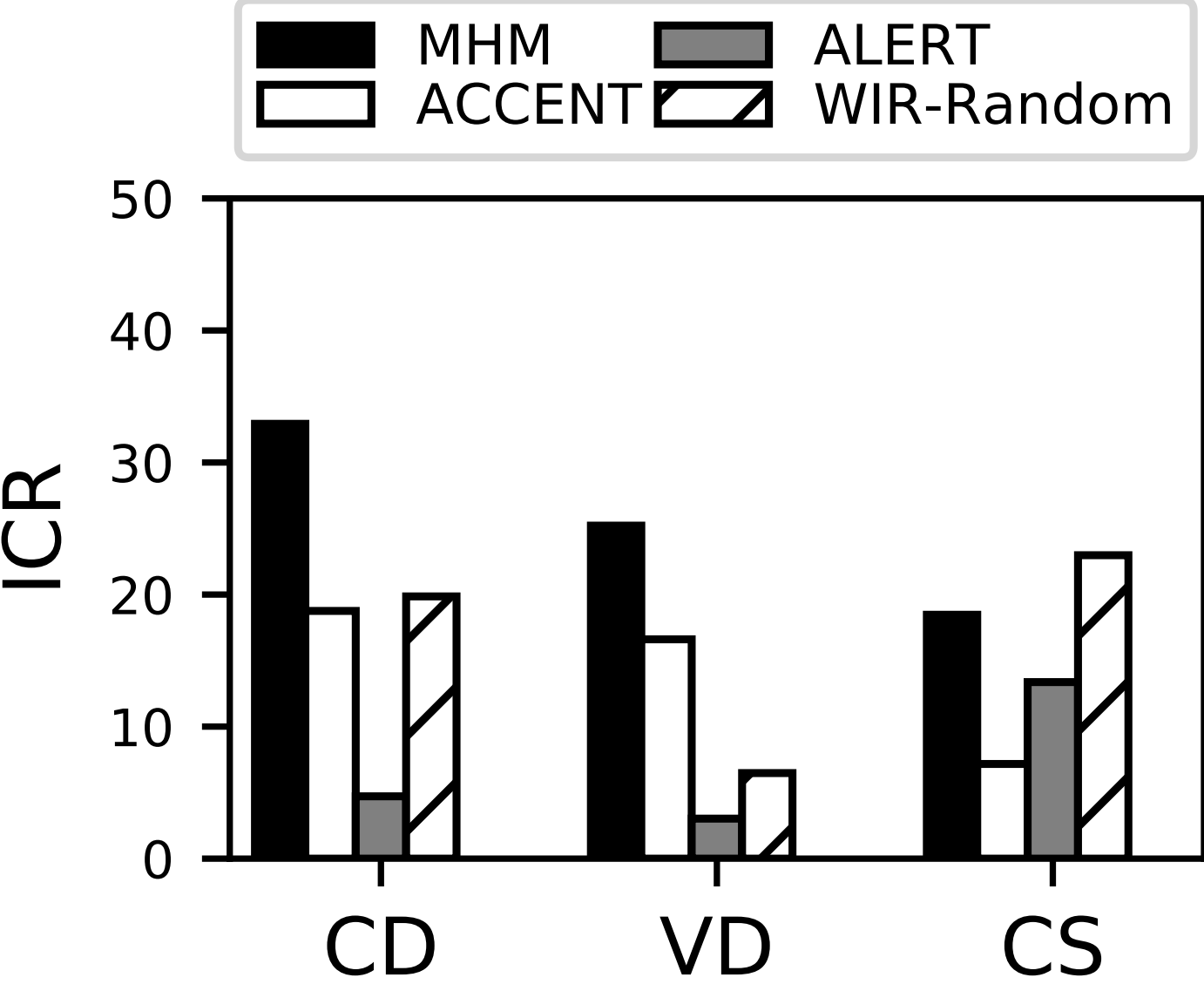}
  \caption{ICR-CodeGPT} \label{fig:ICR_CodeGPT}
\end{subfigure}
  \begin{subfigure}{0.15\textwidth}
  \centering 
 \includegraphics[width=1\linewidth]{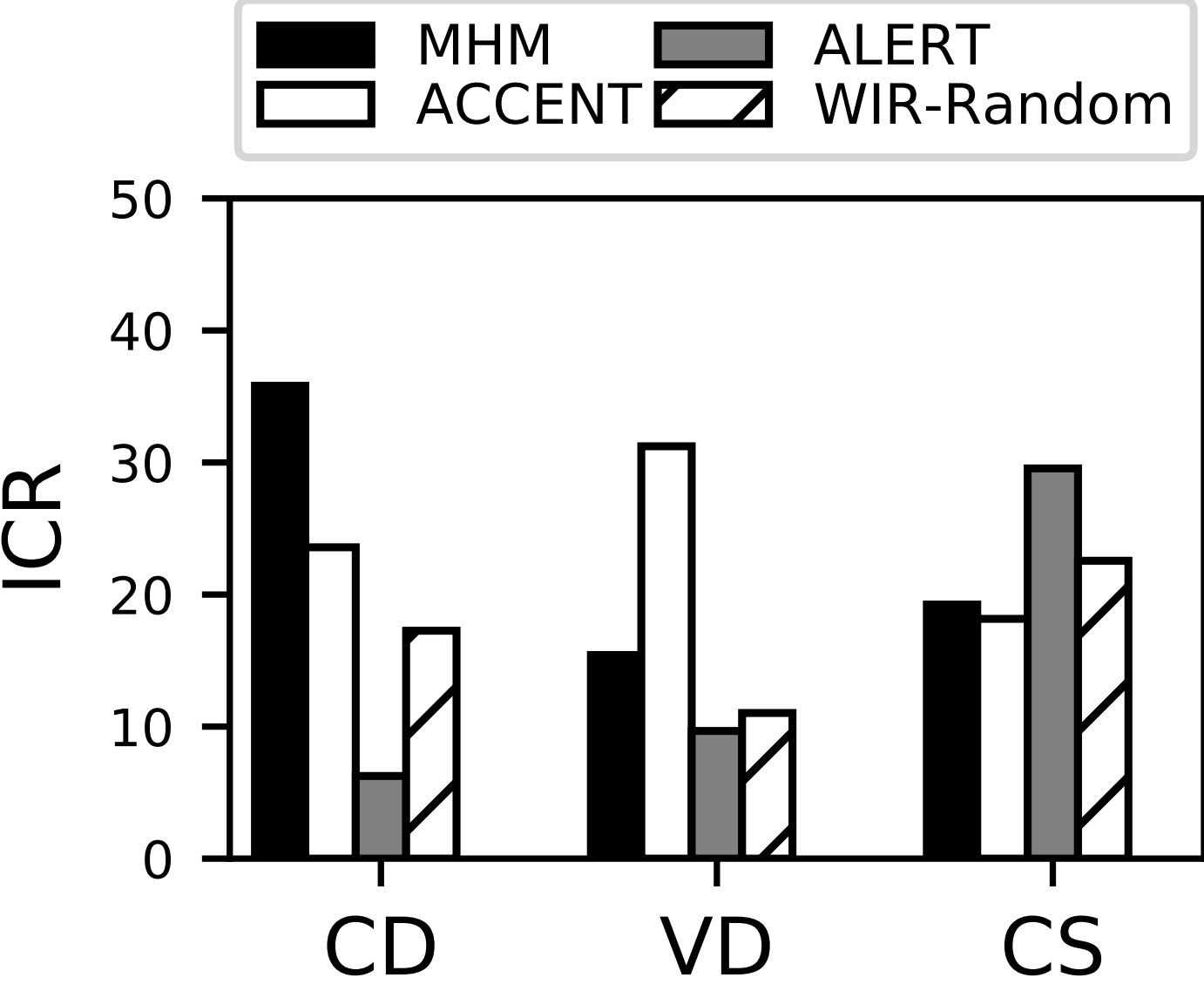}
  \caption{ICR-PLBART}\label{fig:ICR_PLBART}
\end{subfigure}

\begin{subfigure}{0.15\textwidth}  
  \centering
    \includegraphics[width=1\linewidth]{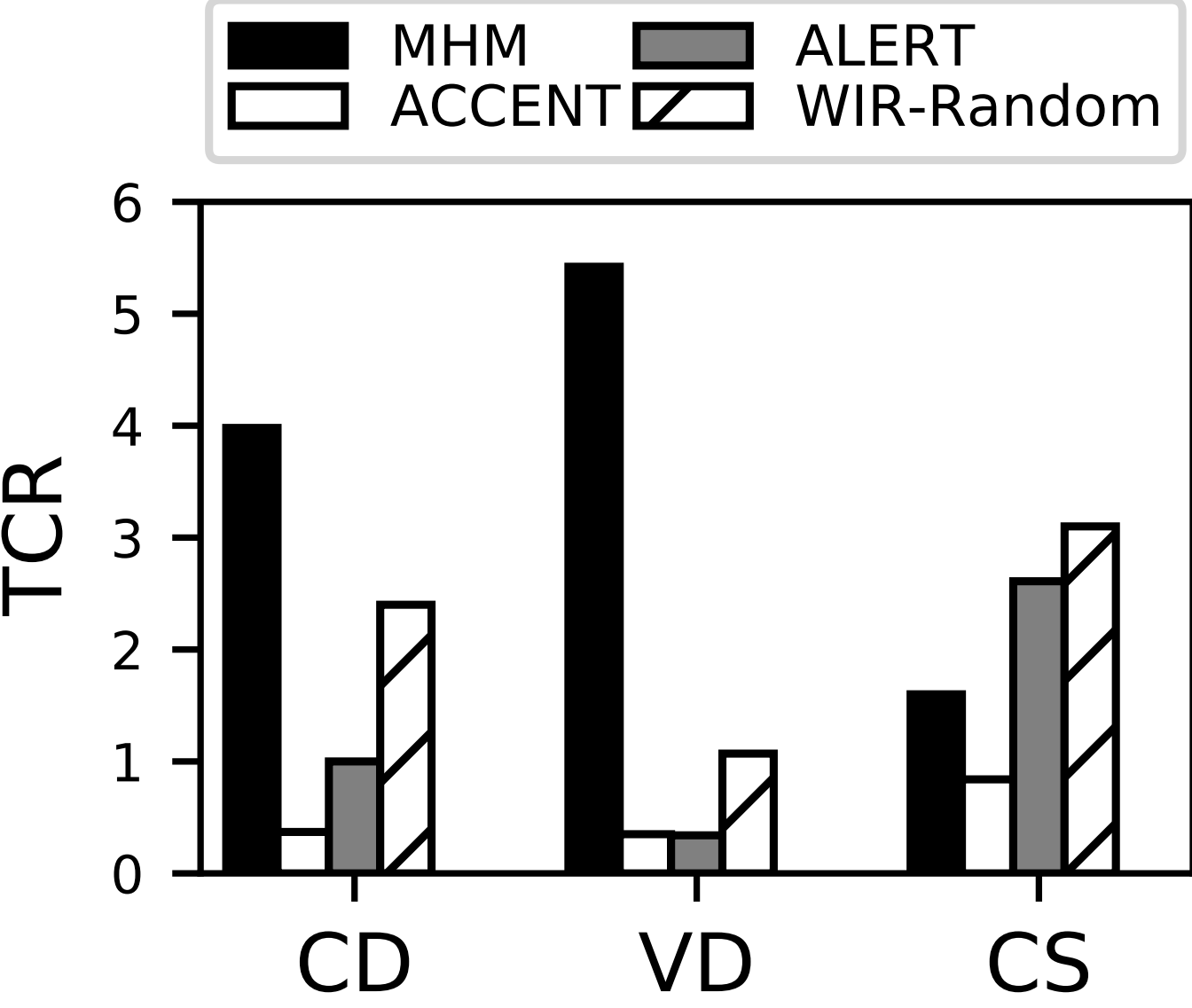}
    \caption{TCR-CodeBERT} \label{fig:TCR_CodeBERT}
\end{subfigure}
\begin{subfigure}{0.15\textwidth} 
  \centering
  \includegraphics[width=1\linewidth]{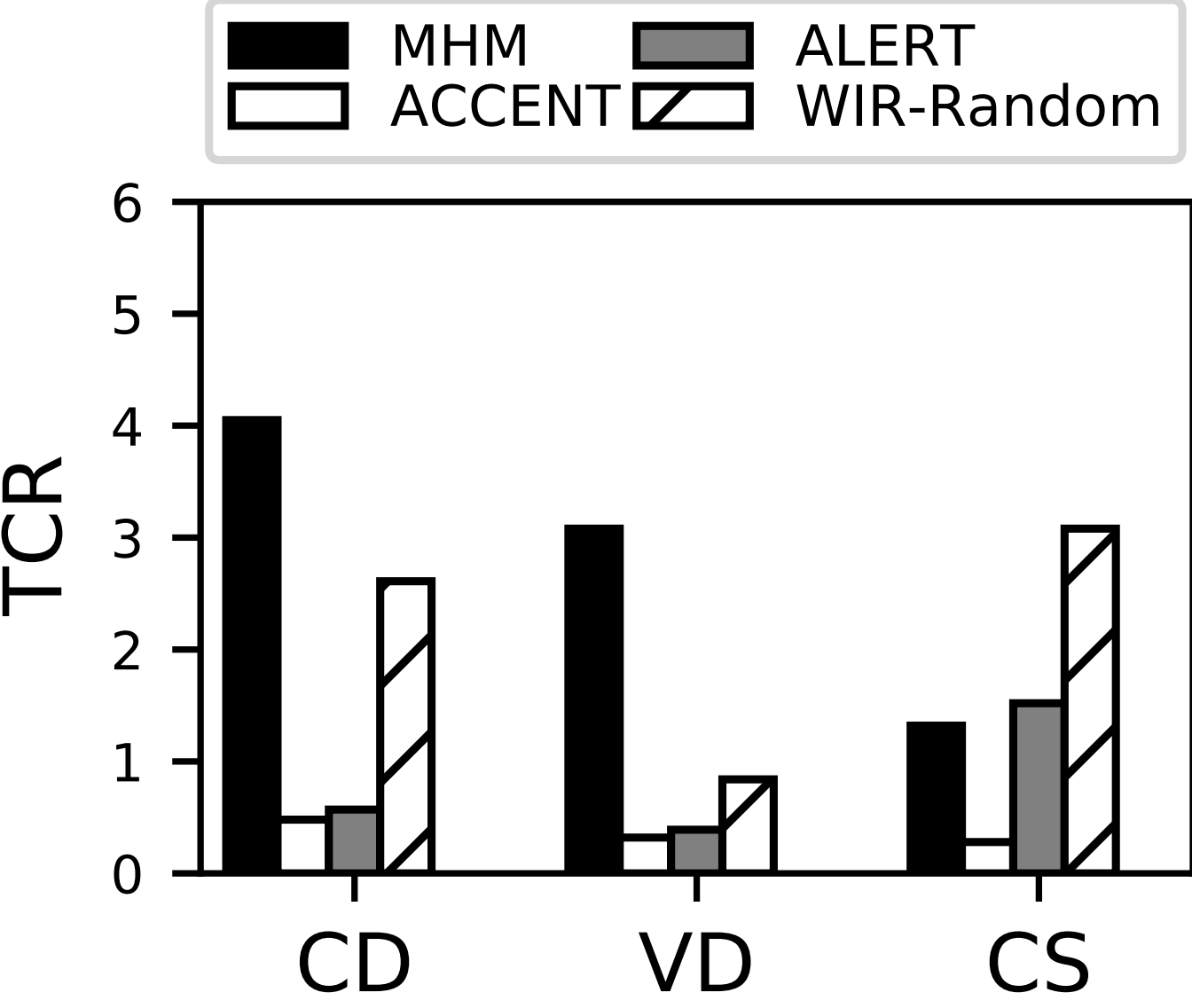}
  \caption{TCR-CodeGPT} \label{fig:TCR_CodeGPT}
\end{subfigure}
  \begin{subfigure}{0.15\textwidth}
  \centering
 \includegraphics[width=1\linewidth]{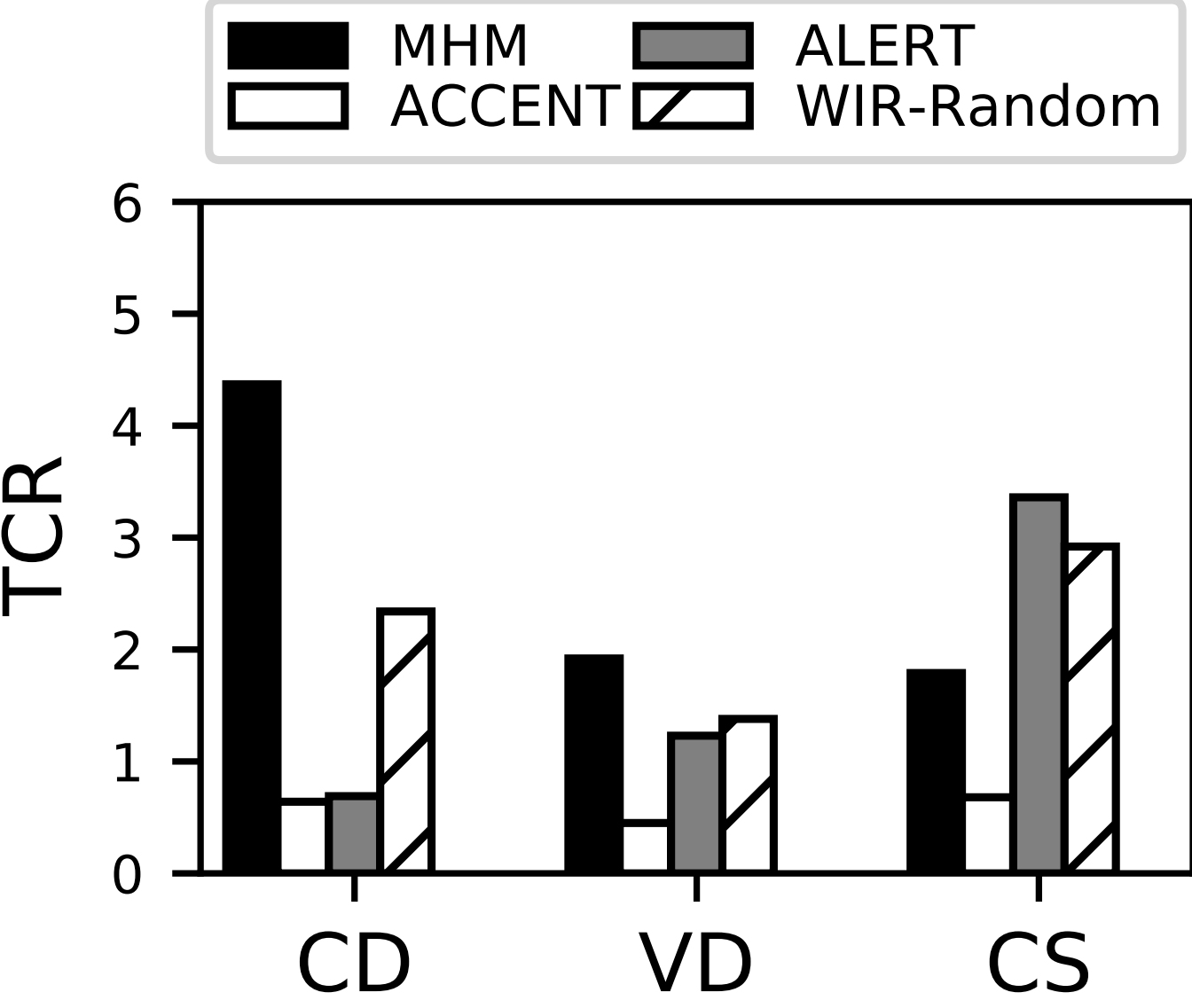}
  \caption{TCR-PLBART}\label{fig:TCR_PLBART}
\end{subfigure}
\begin{subfigure}{0.15\textwidth}
      \centering   
      \includegraphics[width=1\linewidth]{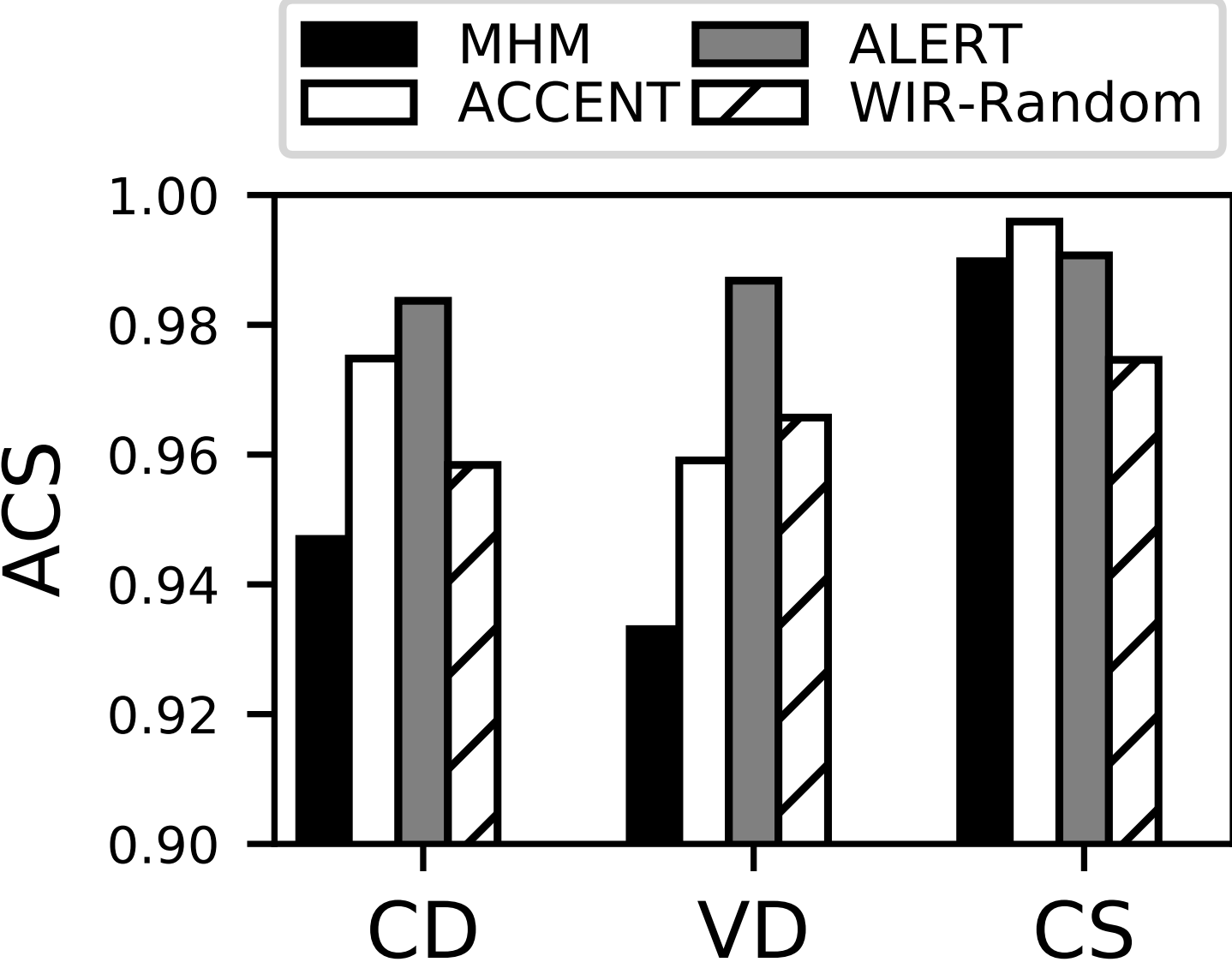}
        \caption{ACS-CodeBERT} \label{fig:ACS_CodeBERT}
    \end{subfigure}   
    \begin{subfigure}{0.15\textwidth}
      \centering   
      \includegraphics[width=\linewidth]{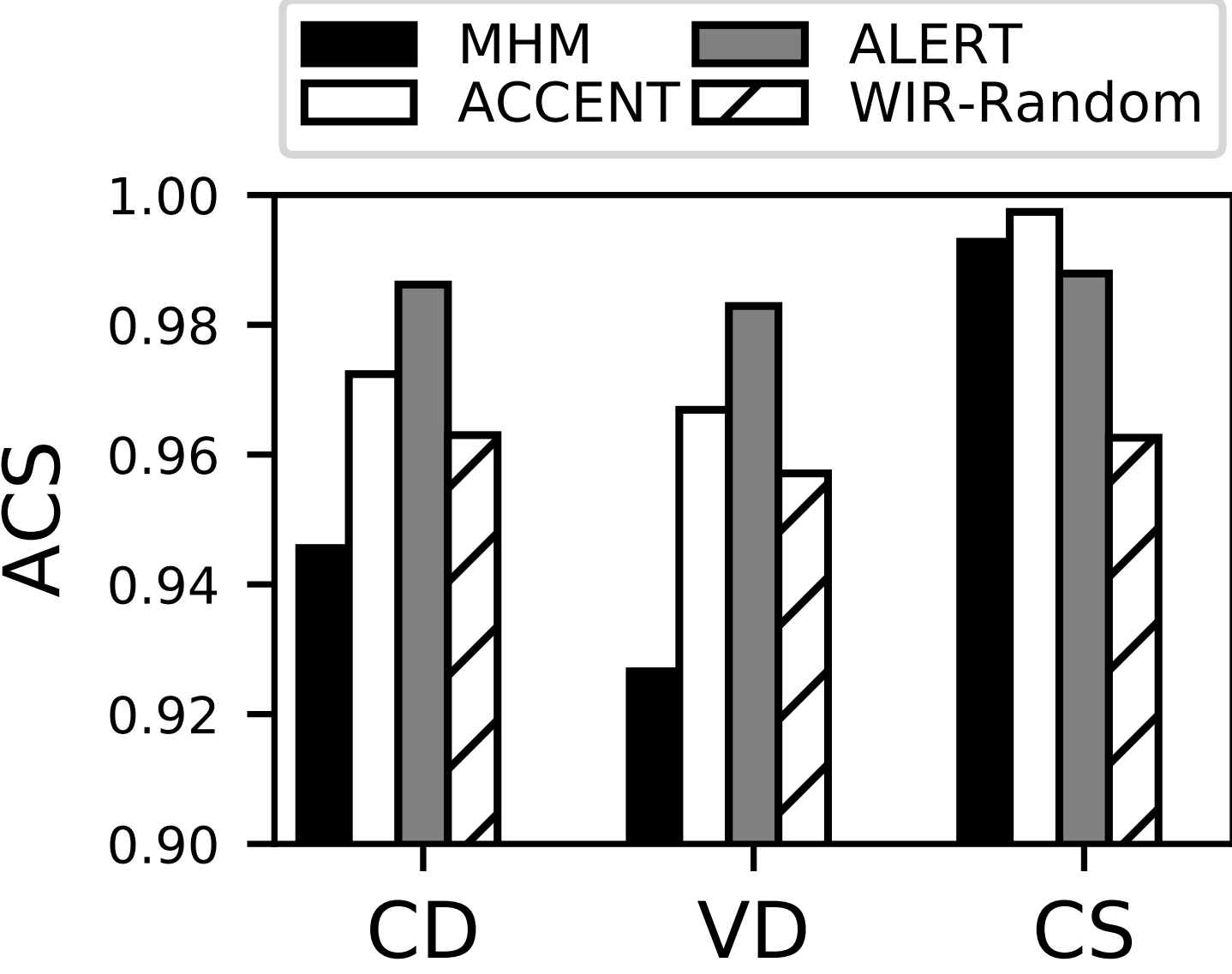}
        \caption{ACS-CodeGPT} \label{fig:ACS_CodeGPT}
    \end{subfigure}
    \begin{subfigure}{0.15\textwidth}
      \centering   
      \includegraphics[width=\linewidth]{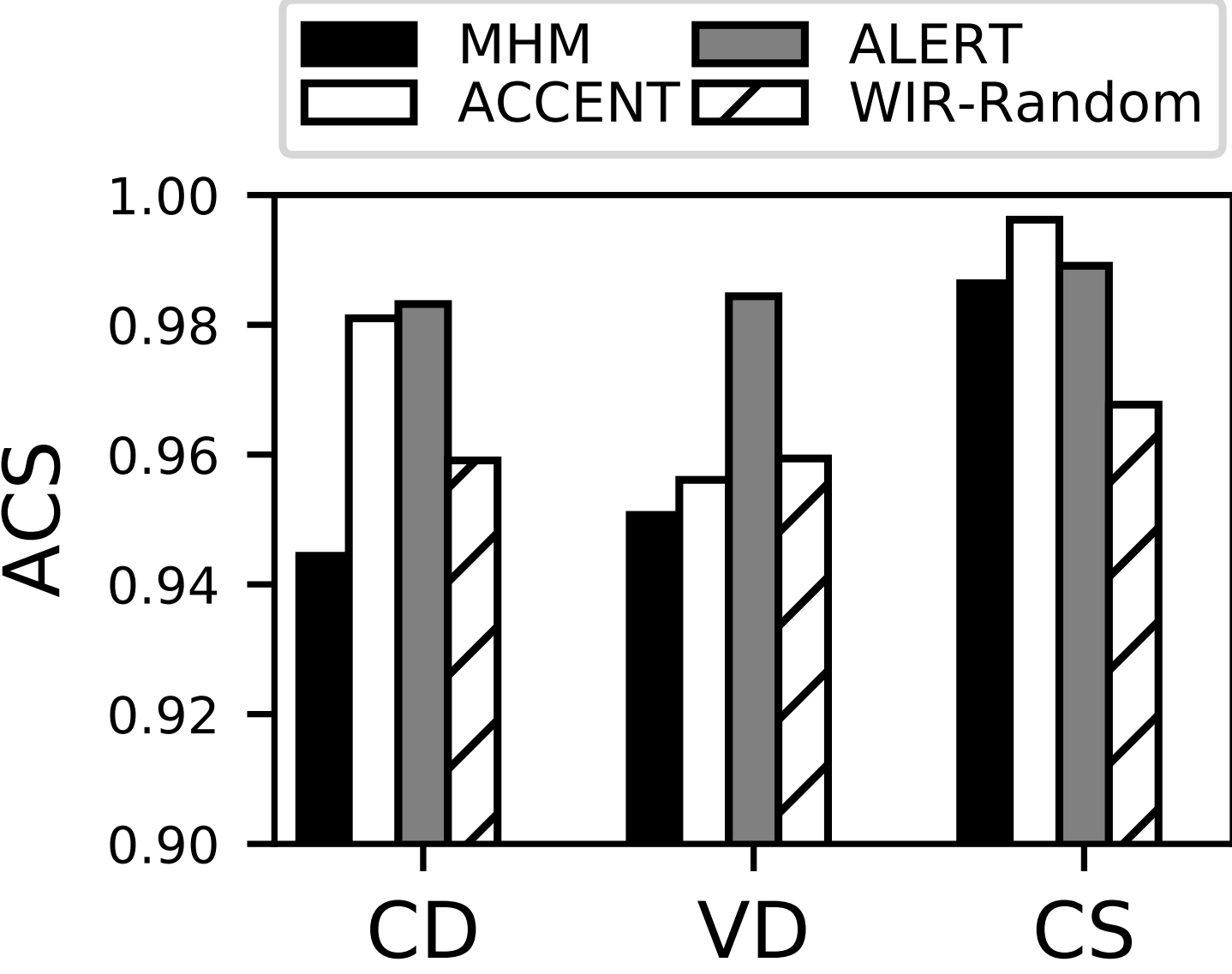}
        \caption{ACS-PLBART} \label{fig:ACS_PLBART}
    \end{subfigure}
    
     \begin{subfigure}{0.15\textwidth}  
   		\centering
     	\includegraphics[width=1\linewidth]{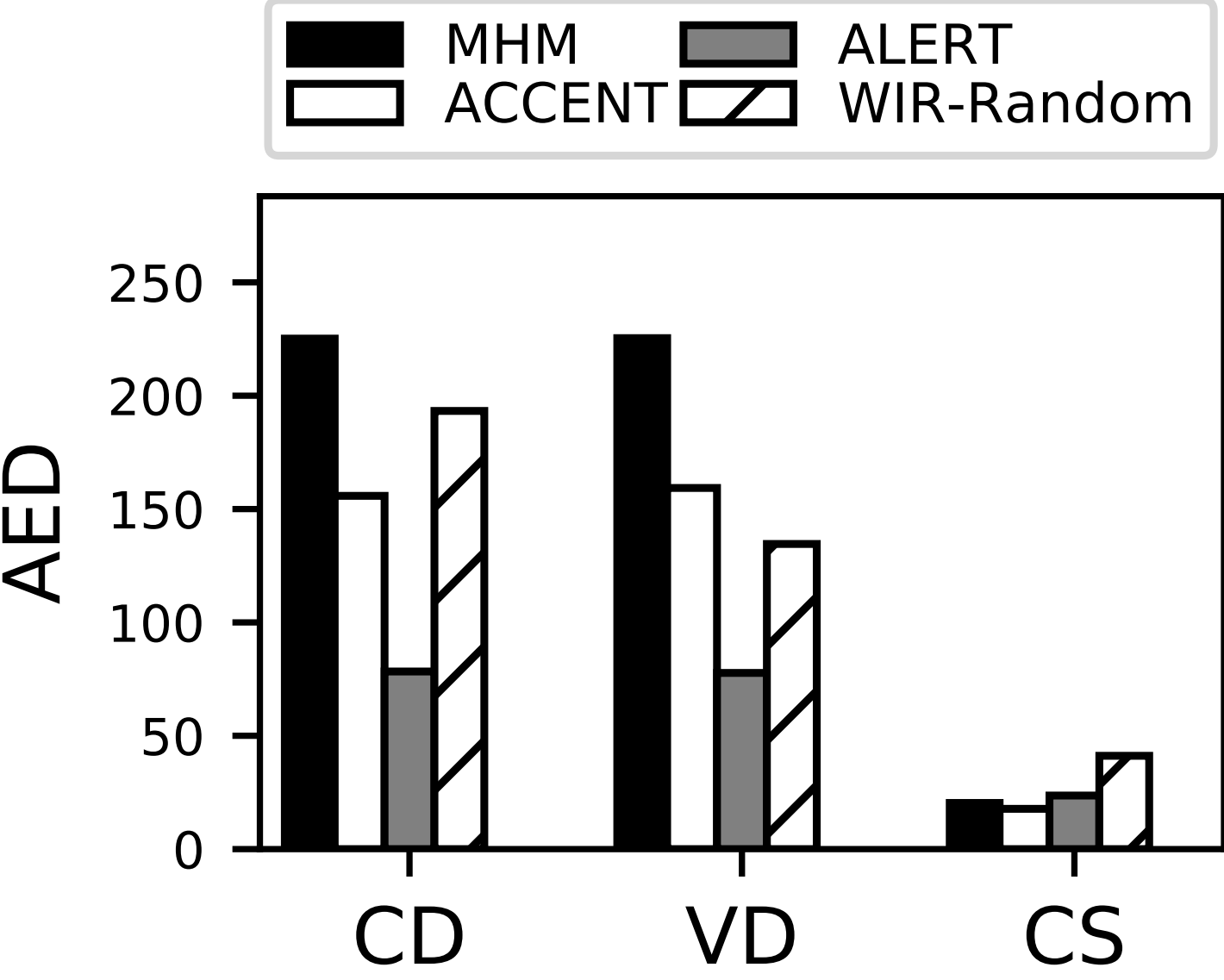}
     		\caption{AED-CodeBERT} \label{fig:AED_CodeBERT}
	\end{subfigure}
	\begin{subfigure}{0.15\textwidth} 
   		\centering
   		\includegraphics[width=\linewidth]{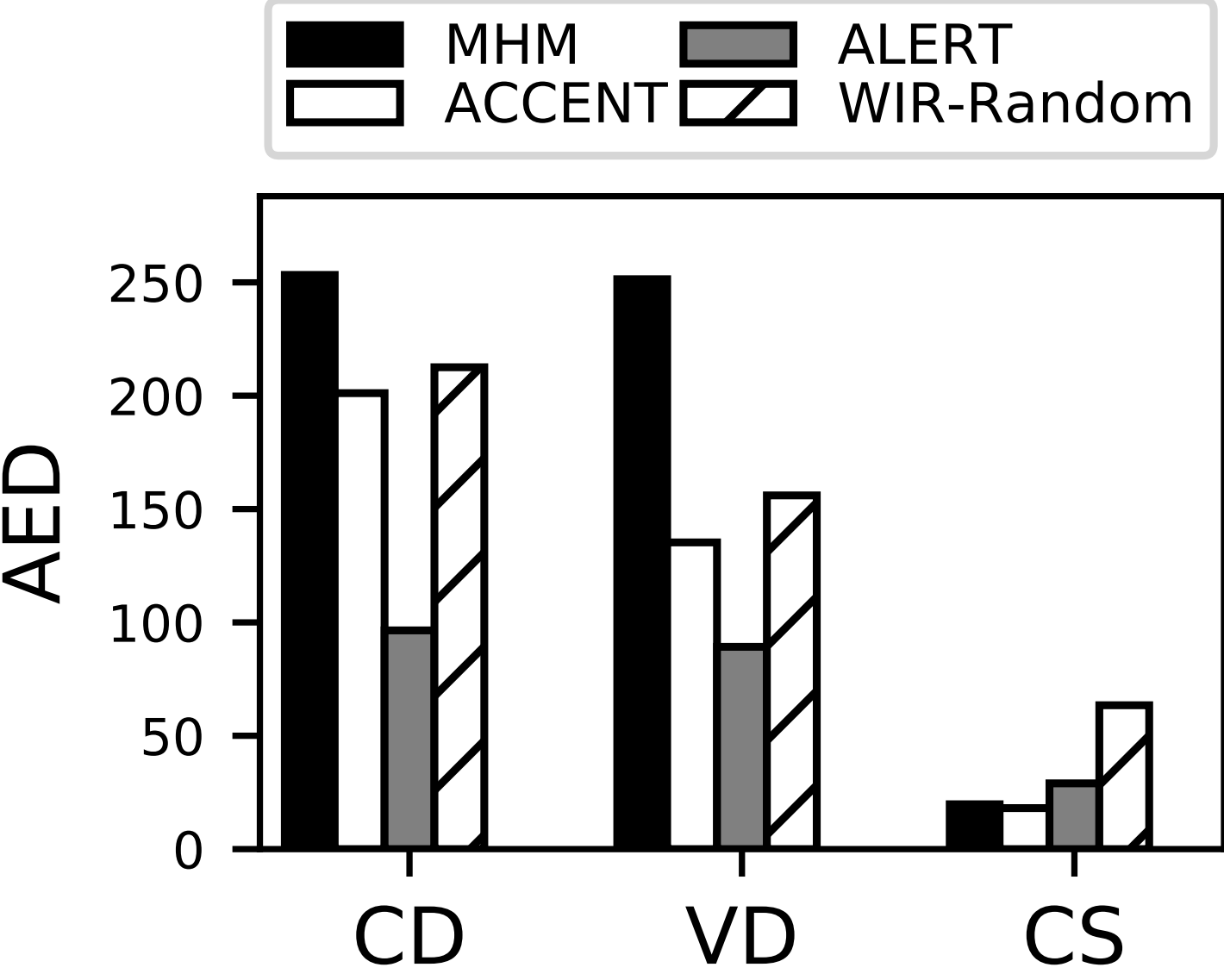}
   		\caption{AED-CodeGPT} \label{fig:AED_CodeGPT}
	\end{subfigure}
   	\begin{subfigure}{0.15\textwidth}
   		\centering
  		\includegraphics[width=\linewidth]{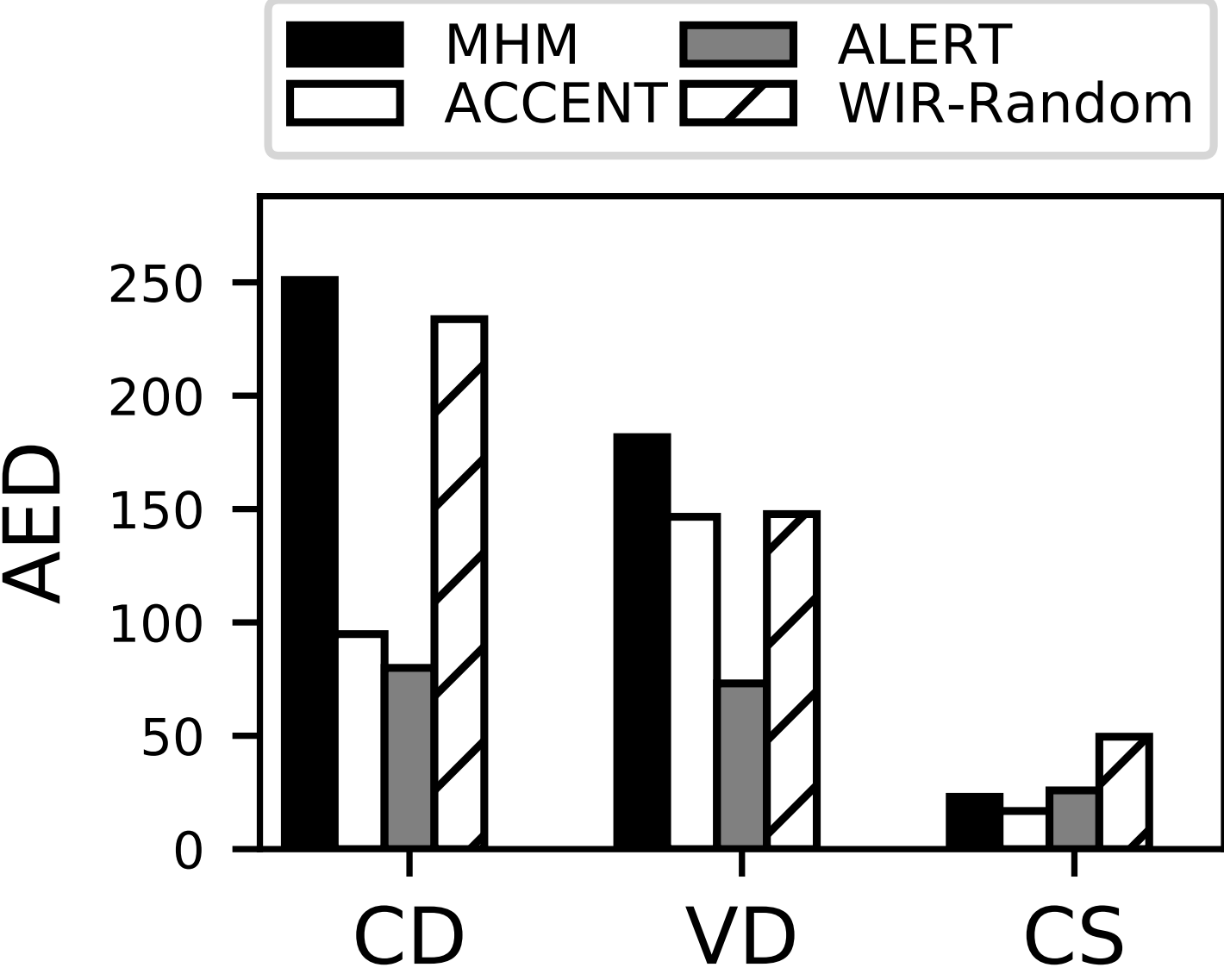}
   		\caption{AED-PLBART}
   		\label{fig:AED_PLBART}
	\end{subfigure}
\vspace{-4mm}
\caption{
\label{fig:ICR_TCR}
Comparison of ICR, TCR, ACS, and AED on attacking CodeBERT, CodeGPT, and PLBART. The lower ICR, AED, and TCR with the higher ACS indicates better performance.
}
\label{fig:boxplot1}
\vspace{-6mm}
 \end{figure}

The above RQ demonstrates the effectiveness and efficiency of adversarial attacks against different PTMCs under various tasks. 
However, with the recent focus on the naturalness of the generated adversarial examples~\cite{DBLP:conf/icse/YangSH022}, a question naturally arises: which attack generates adversarial examples of higher qualities? 
Via analyzing the results of ICR, TCR, ACS and AED are shown in Figure~\ref{fig:ICR_TCR}, we make the following observations.

First, \textit{none of the attacks can achieve the optimal performance on the three tasks in terms of naturalness.}
Specifically, ALERT achieves the best in terms of ICR, ACS, and AED on average, while ACCENT is the  optimal on average against TCR. 
The results are also different on various tasks.
For instance, ALERT outperforms ACCENT on average against ICR, ACS, and AED on clone detection and vulnerability detection, but vice versa on code summarization. 
In general, ACCENT and ALERT outperform MHM and WIR-Random in terms of the averaged ACS, and AED on all tasks since they both consider the naturalness of adversarial examples when replacing identifiers. 

Second, \textit{effective attacks in general generate less natural adversarial examples.}  The adversarial examples generated by MHM and WIR-Random with the highest ASR have the lowest ACS to the original code and the largest ICR, TCR, and AED, indicating that their adversarial code quality is generally lower.
This raises the question towards the usefulness of MHM and WIR-Random in practice since an existing study points out that the adversarial example should not only cheat the model but also be natural to human judges~\cite{DBLP:conf/icse/YangSH022}.
Conversely, the adversarial codes generated by ALERT and ACCENT perform best on both ACS and AED, indicating that these adversarial code are more similar to the original code. Such results also confirm that the adversarial code are more natural than random replacement as claimed by ALERT and ACCENT.
A potential reason for the high attack successful rate of MHM is that less natural perturbations may lead to \textit{Out-of-Distribution} (OOD) examples. 
Such examples could easily lead to the success of adversarial attacks because the models may not perform well on data with different distributions~\cite{DBLP:journals/tmlr/SalehiMHLRS22}.

Third, \textit{the adversarial examples generated by attacking CodeGPT are less similar to the original program than the other pre-trained models.}
Specifically, when the target model is CodeGPT for all attacks, the average ACS is 0.9701, which is lower than 0.9716 and 0.9715 of CodeBERT and PLBART, and the average AED is 127.15, which is higher than 112.72 and 110.35 of CodeBERT and PLBART.
Such differences are all significant as revealed by the Mann-Whitney U test~\cite{MWW} ($p$-value<0.05).
This result also confirms Finding 1, that is, CodeGPT is more resistant to various attacks. Since ASR is negatively correlated with naturalness, the attack algorithm has to choose the suboptimal word with a longer distance among the candidate identifiers to achieve the purpose of misleading CodeGPT.
\begin{center}
\vspace{-2mm}
\begin{tcolorbox}[colback=gray!10, colframe=black, width=8.5cm, arc=1mm, boxrule=0.5pt, 
				  left=1mm, right=1mm, top=0.5mm,bottom=0.5mm]
\textbf{Finding 3: } 
There is a trade-off between the effectiveness and naturalness for adversarial attacks. 
Specifically, effective attacks  generate less natural adversarial examples.
In general, substitution strategies such as context-aware identifier prediction and replacement based on cosine similarity can generate examples of higher qualities than that of random substitution. 

\end{tcolorbox}
\end{center}
\vspace{-5pt}

\vspace{-2mm}
\subsection{Contexts of Perturbed Identifiers (RQ3)}\label{sec:position}
\begin{table*}[t]
    \small
    \caption{Attack success rate for replacing identifiers under different contexts}
    \label{table:position}
    \renewcommand\arraystretch{0.9}
    \setlength\tabcolsep{6pt}
    \centering
    \vspace{-4mm}
    \begin{tabular}{c|ccccc|ccccc|ccccc}
        \toprule
           \multirow{3}{*}{Position} &\multicolumn{5}{c|}{Clone Detection} &\multicolumn{5}{c|}{Vulnerability Detection} & \multicolumn{5}{c}{Code Summarization} \\   \cline{2-16}
                &\multirow{2}{*}{Total}&\multicolumn{2}{c}{MHM} &\multicolumn{2}{c|}{WIR-Random} &\multirow{2}{*}{Total}&\multicolumn{2}{c}{MHM} &\multicolumn{2}{c|}{WIR-Random} &\multirow{2}{*}{Total}&\multicolumn{2}{c}{MHM} &\multicolumn{2}{c}{WIR-Random} \\   \cline{3-6}     \cline{8-11} \cline{13-16}
                & &n=1 &n=3 &n=1 &n=3 & &n=1 &n=3 &n=1 &n=3 & &n=1 &n=3 &n=1 &n=3 \\
        \midrule
        Method  &4,000 &11.01    &17.90  &1.92 &9.91  &4,000 &11.16 &19.81 &2.36 &\textbf{5.70} & 4,000&\textbf{61.32}  &\textbf{88.81} &\textbf{46.84} &\textbf{80.81}  \\  
        Return  &1,650 &11.08     &11.26  &1.83 &2.08  &2,659 &10.60 &10.60 &2.49 &2.49 & 2,303&26.81 & 29.77 &14.22 &18.25  \\  
        If      &2,574 &15.07    &23.37  &3.00 &11.42 &2,773 &\textbf{12.38} &\textbf{21.07} &2.33 &5.47 & 2,446&24.81 &32.40 &13.72 &22.33 \\  
        Throw   &899  &10.81    &13.39  &1.23 &3.07 &2,337 &7.37 &7.37 &1.58 &1.58 &  462&25.21 &27.12 &12.33 &14.25 \\  
        Try     &2,858 &13.78    &23.89  &2.30 &12.77 &2,501 &8.41 &15.24 &1.93 &4.33 &  603&22.87 &30.77 &11.13 &20.85 \\ 
        For     &741  &\textbf{16.11}    &\textbf{26.05}  &\textbf{3.08} &\textbf{14.29} &398 &10.98 &18.40 &\textbf{2.97} &5.64 &  398&26.73 &35.14 &15.32 &29.13\\  
        \bottomrule                                              
    \end{tabular}
    \vspace{-4mm}
\end{table*}
The exploration of the above two RQs demonstrates that there are two factors affecting the effectiveness of adversarial attack, the search algorithm and the identifier substitution strategy. 
However, they both concentrate on how to change the programs, while another important perspective for adversarial attack is determining \textit{\textbf{what}} identifiers should be changed. In this RQ, we investigate if perturbing identifiers under different contexts can cast significant impact on the attacking effectiveness.
Via analyzing the results as shown in Table~\ref{table:position}, we make the following observations.  

\textit{A large number of instances can be successfully attacked even if only one identifier is replaced.} 
Specifically, we limit the replaced identifiers to 1 and 3 respectively.  
The results show that existing techniques can still attack various models successfully, and the ASR increases as such a threshold increases. 
For different attack approaches, MHM is consistent with previous experiments in that ASR is higher than WIR-Random in all cases. 
In this paper, we explore the impact of identifiers in different statements on the models' performance.
Next, we take MHM as an example to make such explorations since the ASR of MHM and WIR-Random on different statements share similar trend.
To ease for presentation, we refer to the different statements by their name, such as \textit{For}.

The results show that \textit{the attacking effectiveness is sensitive to the identifiers under various types of statements, and such sensitivity diverges across various tasks}.
Specifically, in clone detection, replacing identifiers in \textit{For}, \textit{If}, and 
\textit{Try} is more likely to result in a successful attack. 
For instance, when n=3, the ASR of these three statements all exceed 20\%, and the highest is 26.05\% of \textit{For}.
The perturbations to \textit{Method}, \textit{Throw}, and \textit{Return} achieve relatively low ASRs (less  than 20\%), the lowest of which is 11.26\% of \textit{Return}.
In vulnerability detection, the first three most effective statements are \textit{If}, \textit{Method}, and 
\textit{For}.
When n=3, the ASR of these three statements are close to 20\%, and the highest is 21.07\% of \textit{If}.
\textit{Throw} has the lowest impact (ASR=7.37\%).
In code summarization, \textit{Method} dominates the attack effectiveness with an extremely high  ASR of 88.81\%, which is significantly higher than those of the other statements.
Specifically, the ASRs of the remaining types of statements are all below 50\%.
Among them, \textit{Throw} has the lowest impact on the code summarization model, with its ASR reaching 27.12\%.
We further perform a case study to analyze why prioritizing statements can significantly affect the performance of adversarial attacks (see Section~\ref{sec:case_study}).
\begin{center}
\vspace{-2mm}
\begin{tcolorbox}[colback=gray!10, colframe=black, width=8.5cm, arc=1mm, boxrule=0.5pt, 
				  left=1mm, right=1mm, top=0.0mm,bottom=0.5mm]
\textbf{Finding 4: }
The context of the identifiers (\eg~where the identifiers reside) can affect the attacking effectiveness significantly, which suggests that the perturbation strategies should consider the context of identifiers aiming for more effective attacks. 
\end{tcolorbox}
\end{center}
        \vspace{-2mm}
\section{New Approach}\label{sec:approach}
Our empirical investigation reveals two main challenges for adversarial attack against PTMC, which are the trade-off between effectiveness and efficiency (Finding 2) as well as that between effectiveness and naturalness (Finding 3). 
Both the two challenges may compromise the practical usefulness of existing adversarial attacks. 
Aiming to alleviate the second challenge, the state-of-the art approach, ALERT~\cite{DBLP:conf/icse/YangSH022}, adopts a context-aware identifier substitution strategy to improve adversarial code naturalness.
However, our experiment reveals that both the effectiveness and efficiency of ALERT are still limited. For instance, it only achieves an average ASR of 20.32\% on the three tasks. 

Our tool aims to enhance both the effectiveness and efficiency while guaranteeing the naturalness, and the novelty of which is mainly embodied in the following two aspects. 
First, Finding 4 shows that perturbations on different types of statements can achieve varying success rates on existing attack techniques. 
As such, we propose to incorporate such prior knowledge to prioritize identifier selection, thus enhancing the effectiveness and efficiency of the attack.
This attack strategy, which incorporates code features, is different from all the previous works, including the SOTA ALERT.
Second, Finding 2 reveals that the effectiveness and efficiency of existing attacks are still limited. 
The reasons are as follows. 
MHM mainly uses a random method to replace identifiers one by one along the  sequence of identifiers that can be replaced while this strategy requires a large number of queries to mislead the model.
Meanwhile, ALERT, ACCENT, and WIR-Random replace identifiers sequentially. Once the top candidates fall into local optimal solutions, it is difficult for them to find the global optimal solution since they do not repeatedly process the replaced identifiers.
To alleviate such problems, we propose to use \textit{beam search}~\cite{DBLP:journals/ml/KumarVME13} to focus on all the identifiers in a statement, which can simultaneously search from multiple sequences and replace multiple identifiers. 

\vspace{-3mm}
\subsection{Approach Design}
Algorithm~\ref{algorithm:beam} shows the workflow of BeamAttack.
It first obtains the set of identifiers in different statements {\mycode S} and the number of statement types {\mycode T} (Line 2).
The priority of different statements is summarized by our prior knowledge as shown in Table~\ref{table:position}. 
For instance, the prioritized statement types for the clone detection task is: {\mycode For}, {\mycode If}, {\mycode Try}, {\mycode Method}, {\mycode Throw}, {\mycode Return}, and {\mycode Others}.
BeamAttack then performs beam search over different types of statements sequentially to generate new examples.
For the first iteration, the replaced codes are the source program $p$ (Line 3). For subsequent iterations, the replaced code are the $k$ perturbed codes returned by function {\mycode BeamSearch} (Section~\ref{sec:beam}) in the previous iteration. 
The sequence to be replaced (which is denoted as $Seq$) is initialized to all the identifiers in the entire statement (Line 5).
We apply {\mycode BeamSearch} multiple times until the last category of statements is searched or an adversarial example is successfully generated.
Note that we record the replaced identifiers after each {\mycode BeamSearch} (Line 7).
Finally, BeamAttack performs the last {\mycode BeamSearch} with the recorded replaced identifier as $Seq$ (Line 8). 
This step is to alleviate the limitation that beam search will not process those identifiers that have already been replaced.  
Since an identifier can only be replaced by a unique candidate in the adversarial example, some suboptimal candidates are discarded during the search, and they may become optimal after subsequent identifiers are replaced.

\vspace{-2mm}
\subsubsection{BeamSearch}\label{sec:beam}
The maximum iteration in the search is set to be the product of $Seq$'s length and the weight of the statement type (Line 6).
In particular, we set the weight of the most important statement type to 1, and the other weights are set proportionally according to the prior knowledge in Table~\ref{table:position}.
In each iteration in {\mycode BeamSearch}, we apply {\mycode Perturb} (Section~\ref{sec:perturb}) on all the identifiers from the current type of statements. 
After that, {\mycode BeamSearch} selects the k best ones in the current generation and the previous generation to serve for the next iteration. 
Note that the search process will stop if the current iteration fails to generate new qualified candidates.
Since the number of identifiers is proportional to the maximum number of iterations of {\mycode BeamSearch}, it will directly affect the efficiency of BeamAttack.
As shown in Figure~\ref{fig:ASR&Identifier}, the number of identifiers differs for the three tasks.
To balance the attack success rate and efficiency, we set k to 2, 3, and 5 in clone detection, vulnerability detection, and code summarization, respectively.

\vspace{-2mm}
\subsubsection{Perturb}\label{sec:perturb}
{\myfont Perturb} first uses CodeBERT to generate the 30 most similar candidates for an identifier following the four attacks investigated in our study. 
This similarity is based on the cosine similarity between the embeddings from CodeBERT-MLM following ALERT, which is trained with the objective of masked language modeling.  
These candidates are generated in real-time for each {\mycode Perturb}. 
Since some identifiers are changed during the attack, the top 30 candidates for the identifier to be replaced will also change accordingly.
Then, {\myfont Perturb} chooses the identifier in the candidate list that reduces the current probability of true label the most for replacement to guarantee the naturalness.
If the drop in probability changes the model's predicted label or makes the code summary completely independent of the ground truth (\ie BLEU = 0), we consider the attack successful and output the adversarial example and the replaced identifier.
Otherwise, {\myfont Perturb} returns the perturbed code, original identifier, replaced identifier, and the probability for the corresponding replacement. 

\setlength{\textfloatsep}{0pt}
\begin{algorithm}[t!]
\scriptsize
    \algsetup{linenosize=\scriptsize}
    \caption{The Main Workflow of BeamAttack}
    \label{algorithm:beam}
    \LinesNumbered
    \KwIn{
       {\small  source~program~$p$,~beam~size~$k$,~statement~weight~$SW$}\\
    }
    \KwOut{adversarial example $adv$}
    $rv = []$  \ \ \ \# replaced variable \\
    $S, T$ = \textbf{\textit{GetStatements}}(\textit{p})  \tcc{S is the set of identifiers in different statements, T is the number of statement types}  

    $\mathcal{P}^0$ = \{$p, S[0]$\}   \\
    \For{$t = 0\rightarrow T$}{
        ${Seq^t}=S[t]$     \\
        $max\_iter=Length(Seq^t)*SW[t]$
        $\mathcal{P}^{t+1}  \gets \textsc{BeamSearch}(\mathcal{P}^t,  max\_iter)$ \\
        $rv$.append($\mathcal{P}^{t+1}$. $replacedVariable$)\ \\
    }
    $\textsc{BeamSearch}(\{Code^{T}, rv\},  Length(rv))$
    \SetKwFunction{FMain}{\emph{\textsc{BeamSearch}}}
		\SetKwProg{Fn}{Function}{}{}
		\Fn{\FMain{${\mathcal{P}^t}=\{Code^{t}, Seq^t\}, max\_iter$}}{
            \While{not exceed $max\_iter$}{
                $\mathcal{P}^{copy}  \gets \mathcal{P}^{t}$    \\
              \For{$Code_i^{t}$ \rm{in} $\mathcal{P}^{t}$}{
                \For{$Seq_j^{t}$ \rm{in} $\mathcal{P}^{t}$}{
                ${\mathcal{P}_i^{t+1}} \gets  \textbf{\textit{Perturb}}(Code_i^{t}, Seq_j^{t})$    \\
              }
              }
              $\mathcal{P}^{t}  \gets \textbf{\textit{Selection}}(\mathcal{P}^t\cup\mathcal{P}^{t+1}, k)$        \\
              \If{$\mathcal{P}^{t} == \mathcal{P}^{copy}$}{
            \Return $\mathcal{P}^{t}$
        }
        }
        \Return $\mathcal{P}^{t}$
		
  }
\end{algorithm}

\vspace{-2mm}
\subsection{Evaluation}
\subsubsection{Evaluation Datasets}
To evaluate our method thoroughly, we first evaluate it on the dataset as listed in Section~\ref{sec:dataset}. Furthermore, to demonstrate its generalizability, we perform additional evaluations on three other datasets.
For clone detection, we use the filtered \textit{Google Code Jam} (GCJ)~\cite{DBLP:conf/sigsoft/ZhaoH18,DBLP:conf/wcre/WangLM0J20} dataset consisting of 90,102 examples for training and 4,000 for validation and testing respectively.
For vulnerability detection, we use the Juliet Test Suite\footnote{\url{https://samate.nist.gov/SARD/test-suites}}, which is another widely used open source security benchmark~\cite{DBLP:journals/toplas/SpotoBEFLMS19, DBLP:conf/ccs/YeZH14} besides OWASP.
In particular, we utilize the Java sub-dataset and exclude instances with identical function bodies, and finally obtain the training, validation, and testing sets consisting of 23,636, 2,954, and 2,954 examples respectively.
For code summarization, we use TL-CodeSum (TLC)~\cite{DBLP:conf/ijcai/HuLXLLJ18}, which is widely used as a benchmark~\cite{DBLP:conf/icse/ShiWD0H00S22,DBLP:conf/acl/AhmadCRC20}. Similar to Juliet, we filter out duplicate examples and obtain the training, validation, and testing sets consisting of 69,633, 8,700, and 6,445 examples respectively.
We also try to reproduce the results on these datasets, and the results reflect that our fine-tuned models can also achieve similar performance.
\vspace{-2mm}
\subsubsection{Evaluation Results}
We evaluate BeamAttack and ALERT on 18 sets of experiments
(3 pre-trained models × 3 tasks × 2 datasets) as they both use the context-aware identifier prediction as the substitution strategy to guarantee the naturalness of adversarial examples. 
\begin{table*}[!htb]
    \footnotesize
    \caption{Performance comparison between BeamAttack and ALERT}
    \vspace{-4mm}
    \label{table:Beam}
    \renewcommand\arraystretch{1.0}
    \setlength\tabcolsep{1.2pt}
    \begin{tabular}{cc|ccccccc|ccccccc|ccccccc}
    \toprule
    \multicolumn{2}{c|}{\multirow{2}{*}{ \begin{footnotesize}
Attack Results
\end{footnotesize}}} &  \multicolumn{7}{c|}{\begin{footnotesize}
Clone Detection-BCB
\end{footnotesize}}  & \multicolumn{7}{c|}{\begin{footnotesize}
Vulnerability Detection-OWASP
\end{footnotesize}} & \multicolumn{7}{c}{\begin{footnotesize}
Code Summarization-CSN
\end{footnotesize}} \\ \cmidrule(r){3-23} 
    & &ASR\textcolor{red}{$\uparrow$} &AMQ\textcolor{blue}{$\downarrow$} &ART\textcolor{blue}{$\downarrow$} &ICR\textcolor{blue}{$\downarrow$} &TCR\textcolor{blue}{$\downarrow$} &ACS\textcolor{red}{$\uparrow$} &AED\textcolor{blue}{$\downarrow$} &ASR\textcolor{red}{$\uparrow$} &AMQ\textcolor{blue}{$\downarrow$} &ART\textcolor{blue}{$\downarrow$} &ICR\textcolor{blue}{$\downarrow$} &TCR\textcolor{blue}{$\downarrow$} &ACS\textcolor{red}{$\uparrow$} &AED\textcolor{blue}{$\downarrow$} &ASR\textcolor{red}{$\uparrow$} &AMQ\textcolor{blue}{$\downarrow$} &ART\textcolor{blue}{$\downarrow$} &ICR\textcolor{blue}{$\downarrow$} &TCR\textcolor{blue}{$\downarrow$} &ACS\textcolor{red}{$\uparrow$} &AED\textcolor{blue}{$\downarrow$} \\
    \midrule
    
    \multirow{2}{*}{\thead{CodeBERT}} 
    & BeamAttack 	&\textbf{22.43}    & \cellcolor{lightgray!60}\textbf{1,992.97} &\cellcolor{lightgray!60}\textbf{2.69} &9.06   &1.49 &\cellcolor{lightgray!60}\textbf{0.9872} &\cellcolor{lightgray!60}\textbf{66.45}
                    &\cellcolor{lightgray!60}\textbf{5.04}     &\cellcolor{lightgray!60}\textbf{1,956.58}  &2.94 &2.90   &0.43 &0.9815 &94.06
                    &\cellcolor{lightgray!60}\textbf{51.46}    &\cellcolor{lightgray!60}\textbf{371.82}  &\cellcolor{lightgray!60}\textbf{1.53} &\cellcolor{lightgray!60}\textbf{15.05}   &\cellcolor{lightgray!60}\textbf{1.80} &\cellcolor{lightgray!60}\textbf{0.9951} &\cellcolor{lightgray!60}\textbf{15.91}      \\
    & ALERT  	    &14.48    &2,263.53 &4.49 &\cellcolor{lightgray!60}\textbf{8.77}   &\cellcolor{lightgray!60}\textbf{1.00} &0.9837 &78.35
                    &4.23     &2,718.57  &\cellcolor{lightgray!60}\textbf{2.02} &\cellcolor{lightgray!60}\textbf{2.79}   &\textbf{0.34} &\cellcolor{lightgray!60}\textbf{0.9868} &\cellcolor{lightgray!60}\textbf{77.72}       
                    &48.58    &565.63  &2.30 &23.68   &2.61 &0.9907 &23.63 \\
    \multirow{2}{*}{\thead{CodeGPT}} 
    & BeamAttack 	&\cellcolor{lightgray!60}\textbf{12.09}    &\cellcolor{lightgray!60}\textbf{2,364.90} &\cellcolor{lightgray!60}\textbf{3.40} &7.42   &1.18 &0.9800 &106.91
                    &\cellcolor{lightgray!60}\textbf{6.11}     &\cellcolor{lightgray!60}\textbf{1,904.41} &3.02 &3.50   &0.56 &0.9784 &104.26
                    &\cellcolor{lightgray!60}\textbf{23.86}    &\cellcolor{lightgray!60}\textbf{400.26} &\cellcolor{lightgray!60}\textbf{4.32} &\cellcolor{lightgray!60}\textbf{7.20}   &\cellcolor{lightgray!60}\textbf{0.84} &\cellcolor{lightgray!60}\textbf{0.9948} &\cellcolor{lightgray!60}\textbf{15.71}      \\
                    
    & ALERT  	    &6.59    &2,596.18 &5.89	&\textbf{4.72}   &\cellcolor{lightgray!60}\textbf{0.57} &\cellcolor{lightgray!60}\textbf{0.9862} &\textbf{96.48}

                    &4.75     &2,798.29 &\cellcolor{lightgray!60}\textbf{1.55}  &\textbf{3.03}   &\cellcolor{lightgray!60}\textbf{0.39} &\cellcolor{lightgray!60}\textbf{0.9829} &\cellcolor{lightgray!60}\textbf{89.23}
    
                    &23.50     &938.39 &8.78 &13.37    &1.52 &0.9879 &29.01
 \\    
	\multirow{2}{*}{\thead{PLBART}} 
    & BeamAttack 	&\textbf{13.14}     &\cellcolor{lightgray!60}\textbf{2,327.70}  &\cellcolor{lightgray!60}\textbf{5.75}   &\cellcolor{lightgray!60}\textbf{6.02}  &0.99   &\textbf{0.9856} &\cellcolor{lightgray!60}\textbf{76.76} 
                    &\cellcolor{lightgray!60}\textbf{19.17}      &\cellcolor{lightgray!60}\textbf{1,742.51}   &\cellcolor{lightgray!60}\textbf{3.40}   &\cellcolor{lightgray!60}\textbf{7.05}  &\cellcolor{lightgray!60}\textbf{1.15}   &\cellcolor{lightgray!60}\textbf{0.9861} &\cellcolor{lightgray!60}\textbf{65.79}
                    &\cellcolor{lightgray!60}\textbf{58.29}     &\cellcolor{lightgray!60}\textbf{361.22}   &\cellcolor{lightgray!60}\textbf{1.19}   &\cellcolor{lightgray!60}\textbf{18.21}  &\cellcolor{lightgray!60}\textbf{2.26}   &\cellcolor{lightgray!60}\textbf{0.9948} &\cellcolor{lightgray!60}\textbf{16.29}      \\
    & ALERT  	    &9.10     &2,549.49	&8.39	&6.26	&\cellcolor{lightgray!60}\textbf{0.69}	&0.9832	&79.94
                    &17.28     &2,592.32	&3.65	&9.67	&1.23	&0.9844	&73.10
                    &54.37     &488.82	&1.70	&29.55	&3.36	&0.9891	&25.90 \\     
    \midrule
    \midrule
    \multicolumn{2}{c|}{ \begin{footnotesize}
Attack Results
\end{footnotesize}} &  \multicolumn{7}{c|}{\begin{footnotesize}
Clone Detection-GCJ
\end{footnotesize}}  & \multicolumn{7}{c|}{\begin{footnotesize}
Vulnerability Detection-Juliet
\end{footnotesize}} & \multicolumn{7}{c}{\begin{footnotesize}
Code Summarization-TLC
\end{footnotesize}} \\
    \midrule
	\multirow{2}{*}{\thead{CodeBERT}} 
    & BeamAttack   &\cellcolor{lightgray!60}\textbf{11.84}    &4,702.12 &6.38 &4.52   &1.02 &\cellcolor{lightgray!60}\textbf{0.9795} &\cellcolor{lightgray!60}\textbf{82.00}
                    &\textbf{0.17}     &\cellcolor{lightgray!60}\textbf{1,176.07}  &\cellcolor{lightgray!60}\textbf{0.95} &0.10   &0.01 &\textbf{0.9868} &85.40       
                    &\textbf{44.23}    &602.48  &\cellcolor{lightgray!60}\textbf{4.90} &\cellcolor{lightgray!60}\textbf{13.35}   &2.07 &\cellcolor{lightgray!60}\textbf{0.9924} &\cellcolor{lightgray!60}\textbf{21.52}  	      \\
    & ALERT  	   &6.85    &\cellcolor{lightgray!60}\textbf{4,008.96} &\cellcolor{lightgray!60}\textbf{2.86} &\cellcolor{lightgray!60}\textbf{4.13}   &\cellcolor{lightgray!60}\textbf{0.80} &0.9739 &95.19
                    &0.14     &1,872.35  &1.78 &\textbf{0.05}   &\textbf{0.00} &0.9837 &\textbf{30.50}       
                    &31.87    &\textbf{516.50}  &6.54 &14.13   &\cellcolor{lightgray!60}\textbf{1.98} &0.9877 &26.63  \\  
        \multirow{2}{*}{\thead{CodeGPT}} 
    & BeamAttack    &\cellcolor{lightgray!60}\textbf{25.72}    &\textbf{3,147.57} &5.73 &\cellcolor{lightgray!60}\textbf{6.60}   &1.65 &\cellcolor{lightgray!60}\textbf{0.9861} &\cellcolor{lightgray!60}\textbf{55.97}
                    &\textbf{0.85} &\cellcolor{lightgray!60}\textbf{1,554.57} &\cellcolor{lightgray!60}\textbf{1.21} &0.75   &0.10 &0.9877 & 139.88     
                    &\cellcolor{lightgray!60}\textbf{58.58} &\cellcolor{lightgray!60}\textbf{255.56}  &\cellcolor{lightgray!60}\textbf{1.69} &\cellcolor{lightgray!60}\textbf{13.42}   &\cellcolor{lightgray!60}\textbf{1.76} &\cellcolor{lightgray!60}\textbf{0.9948} &\cellcolor{lightgray!60}\textbf{13.70}   	      \\
    & ALERT  	    &20.42    &3,170.36 &\cellcolor{lightgray!60}\textbf{2.77} &8.70   &\cellcolor{lightgray!60}\textbf{1.53} &0.9813 &72.60
                    &0.27     &1,847.87  &1.92 &\textbf{0.22}   &\textbf{0.02} &\textbf{0.9889} &\textbf{92.00}       
                    &52.66    &603.25  &6.44 &19.90   &2.52 &0.9904 &20.31    \\ 
        \multirow{2}{*}{\thead{PLBART}} 
    & BeamAttack    &\cellcolor{lightgray!60}\textbf{47.32}    &3,446.06 &4.96 &\cellcolor{lightgray!60}\textbf{13.45}   &\cellcolor{lightgray!60}\textbf{3.31} &\cellcolor{lightgray!60}\textbf{0.9851} &\cellcolor{lightgray!60}\textbf{75.85}
                    &\textbf{1.09} &\cellcolor{lightgray!60}\textbf{1,747.10} &2.53 &0.56   &0.07 &0.9913 & 74.62     
                    &\cellcolor{lightgray!60}\textbf{67.29} &\cellcolor{lightgray!60}\textbf{368.27}  &\cellcolor{lightgray!60}\textbf{2.40} &\cellcolor{lightgray!60}\textbf{17.78}   &\cellcolor{lightgray!60}\textbf{2.44} &\cellcolor{lightgray!60}\textbf{0.9945} &\cellcolor{lightgray!60}\textbf{15.15}   	 	     \\
    & ALERT  	    &36.67    &\cellcolor{lightgray!60}\textbf{2,537.58} &\cellcolor{lightgray!60}\textbf{3.20} &19.54   &3.71 &0.9794 &98.24
                    &0.78     &1,973.91  &\textbf{2.28} &\textbf{0.45}   &\textbf{0.04} &\textbf{0.9953} &\textbf{56.48}       
                    &55.27    &521.31  &6.22 &23.79   &3.37 &0.9889 &23.36        \\ 
    
    \bottomrule
    \multicolumn{23}{l}{\footnotesize Note: Bold numbers indicate the better performance for the given metric. The cell with lightgray background denotes the outperformance is significant ($p$ < 0.05).
    }
    \end{tabular}
   \vspace{-2mm}
\end{table*}
The evaluation results are summarized in Table~\ref{table:Beam}, and we can observe that BeamAttack consistently achieves higher attack success rates on all experiments.
On average, BeamAttack outperforms ALERT by 20.85\% in terms of ASR, showing that it is more effective in achieving successful attacks.
With respect to the attack efficiency, BeamAttack outperforms ALERT on 15/18 experiments.
In addition, the average AMQ of BeamAttack is 1,690.12, which is 11.98\% less than that of ALERT (1,920.18 on average).
This indicates that our method, which relies on the statement importance to search for adversarial examples, is more efficient. 
Since BeamAttack replaces identifiers based on context-aware identifier prediction, the adversarial examples generated by it are of higher qualities with lower perturbation rates. 
Specifically, the average ICR of BeamAttack is 8.49, which is lower than that of ALERT (11.32 on average).
Meanwhile, the average ICR and TCR of BeamAttack on the 18 experiments are 8.16 and 1.29, which are lower than ALERT (10.71 and 1.43, respectively).
The average ACS of BeamAttack is 0.9871, which outperforms ALERT (0.9861).
BeamAttack only performs worse than ALERT on AED (62.57 vs. 60.54).

To verify whether the performance differences are statistically significant, we apply the one-sided Mann-Whitney U tests~\cite{MWW} to each experiment. Significant differences ($p$-value<0.05) are marked with lightgray background in Table~\ref{table:Beam}.
The results show that our method outperforms ALERT on 69.05\% of the cases (\ie  87/126, 18 experiments * 7 metrics). 
Among these cases, 78 (accounting for 89.66\%) demonstrate significant differences, strongly verifying that our method not only surpasses ALERT on most evaluation metrics, but also achieves significant outperformance.
Although ALERT performs relatively better in the remaining 39 cases, we note the difference between BeamAttack and ALERT is significant for only 22 cases.
\section{Discussion}
\subsection{Case Study}
\label{sec:case_study}
\begin{figure*}[t]
  \centering

  \begin{subfigure}[b]{0.31\textwidth}
    \centering
    \includegraphics[width=\textwidth]{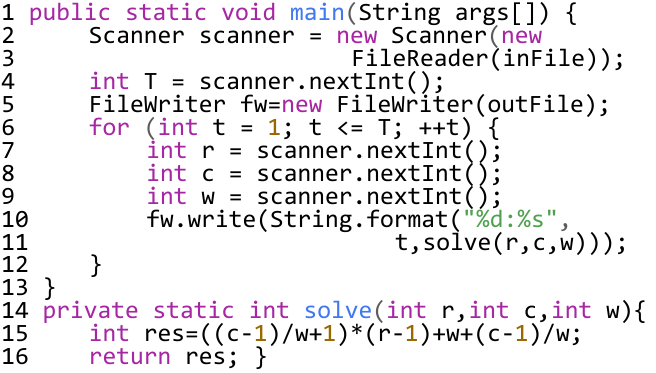}
    \caption{Clone Detection: the first version of index \#1216 from dataset GCJ}
    \label{fig:case_clone}
  \end{subfigure}
  \hfill
  \begin{subfigure}[b]{0.31\textwidth}
    \centering
    \includegraphics[width=\textwidth]{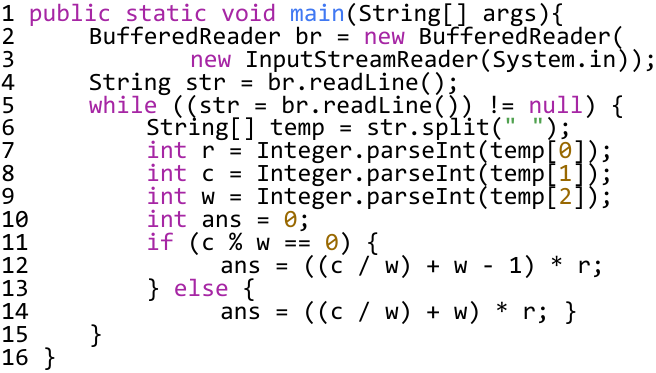}
    \caption{Clone Detection: the cloned version of index \#1216 from dataset GCJ}
    \label{fig:case_clone2}
  \end{subfigure}
  \hfill
  \begin{subfigure}[b]{0.31\textwidth}
    \centering
    \includegraphics[width=\textwidth]{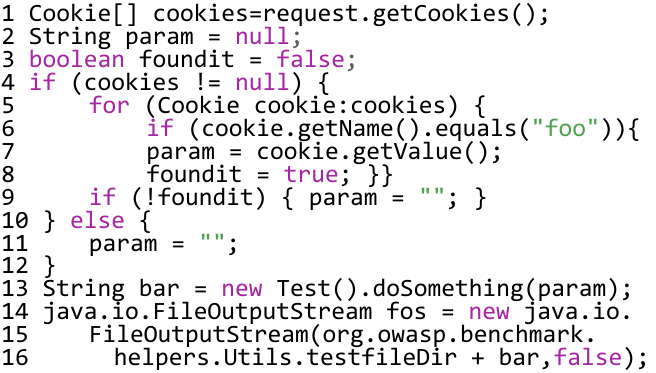}
    \caption{Vulnerability Detection: Index \#26 from dataset OWASP}
    \label{fig:case_vul}
  \end{subfigure}
\vspace{-3mm}
  \caption{Case study on clone detection and vulnerability detection}
  \vspace{-4mm}
  \label{fig:case_study}
\end{figure*}
In this section, we perform an additional case study to qualitatively compare BeamAttack with ALERT to understand their distinctions and why prioritizing statements can significantly affect the performance of adversarial attacks.
We illustrate based on Figure~\ref{fig:case_study} on clone detection and vulnerability detection.
The cloned code corresponding to Figure~\ref{fig:case_clone} transforms {\mycode for} loop into {\mycode while} and refactors the {\mycode solve} function, as shown in Figure~\ref{fig:case_clone2}. 
Figure~\ref{fig:case_vul} displays a code snippet from the OWASP dataset that contains a \textit{Path Traversal} vulnerability on lines 14-16.
The vulnerability exists due to the lack of input validation for the variable {\mycode bar} which receives data from the {\mycode cookie} (i.e., {\mycode param}). Such validation is often performed by using the {\mycode if} statement, which highlights the importance of {\mycode if} in influencing the predictions for vulnerability detection models.
These two cases can reflect our finding in RQ3 that \textit{the statements have the most significant impact on adversarial attacks against clone detection and vulnerability detection are {\mycode For} and {\mycode If}, respectively.}
Specifically, during the attack process, the first set of identifiers extracted by BeamAttack are: [ {\mycode For: [fw, r, c, t, T, w, scanner]}] and [ {\mycode If}: [{\mycode cookie}, {\mycode foundit}, {\mycode i}, {\mycode cookies}, {\mycode bar}, {\mycode param} ]] respectively since it prioritizes those statements based on the learned prior knowledge. 
One successful attack substitution follows [ {\mycode w}$\mapsto${\mycode j}, {\mycode c}$\mapsto${\mycode k}, {\mycode fw}$\mapsto${\mycode ww} ] and [ {\mycode cookies}$\mapsto${\mycode Cooks}, {\mycode param}$\mapsto${\mycode ram}, {\mycode bar}$\mapsto${\mycode ban}, {\mycode i}$\mapsto${\mycode vi}, {\mycode foundit}$\mapsto${\mycode foundait}, {\mycode doSomething}$\mapsto${\mycode runNothing}] respectively.
It can be seen that successful attacks can be achieved by only replacing part of the identifiers in the {\mycode For} and {\mycode If} statements. 
In contrast, the sequences replaced by ALERT are: [ {\mycode c}, {\mycode fw}, {\mycode w}, {\mycode res}, {\mycode r}, {\mycode inFile}, {\mycode outFile}, {\mycode t}, {\mycode T}, {\mycode scanner} ] 
and [ {\mycode bar}, {\mycode param}, {\mycode response}, {\mycode foundit}, {\mycode fos}, {\mycode i} ] respectively. 
We note that ALERT does not accurately replace the identifiers required to achieve the attack. For example, it replaces irrelevant identifiers {\mycode inFile} and {\mycode outFile} in clone detection, and also misses the identifier {\mycode cookies} that is highly relevant to the vulnerability. 
The above analysis explains the distinction between ALERT and BeamAttack as well as  why ALERT is less effective.

In addition to the difference in prioritizing identifier substitution, it is worth noting that we also utilize beam search to attack, which allows more sequences for identifier substitutions within the same statement.
For instance, in Figure~\ref{fig:case_vul}, ALERT replaces identifiers strictly according to a fixed order, meaning only {\mycode param} can come after {\mycode bar}. 
In contrast, after replacing {\mycode bar}, BeamAttack can choose {\mycode foundit}, {\mycode param}, or {\mycode i} as the next replaceable identifier, which mitigates the risk of getting stuck in local solutions.
\vspace{-2mm}
\subsection{Implications}
For model robustness, we find that existing PTMCs are susceptible to adversarial attacks (Finding 1), which presents a considerable challenge to their robustness. 
Therefore, we strongly advocate that researchers need to place equal, if not greater, emphasis on improving model robustness while striving to improve model performance. 
Practical strategies such as adversarial training or data augmentation can be employed to enhance robustness.
These methods can equip models with the resilience required to counter adversarial perturbations and improve their overall reliability.

For adversarial attack approach, considering the trend of \textit{large language models} (LLMs) towards being closed-source and chargeable, attacks through massive queries may become extremely costly or even infeasible. Therefore, a further exploration of the relationship between code identifiers and model prediction results is necessary to derive a more accurate sequence for identifier replacement.
Additionally, {we can leverage the SOTA language generation techniques, such as ChatGPT}, to replace identifiers and enhance the naturalness of adversarial examples.
Finally, considering the effectiveness of statement prioritization, future methods should focus on a more in-depth analysis of code structure, such as employing comprehensive semantic analysis to devise more effective attacks.
        \vspace{-2mm}
\section{Threats to Validity}
\textbf{Internal validity:} Parameter settings such as the number of iterations can lead to different results.
We adopt the following strategy to mitigate this threat.
When the parameters can be set uniformly, we set the parameters consistently with the five attacks, such as the number of candidates for the identifier.
When the parameters are specific to an attack approach, we follow the settings in the original paper exactly to achieve fair comparisons.
Another internal validity threat is the potential bugs in our implementation.
To reduce such threats,
we have carefully checked our implementations and also open sourced all the materials and code to the community for further checks.

\noindent\textbf{External validity:} External validity is threatened by the generalizability of tasks, datasets, and models. 
For tasks, 
the selected ones have been extensively studied in existing works on adversarial attacks~\cite{DBLP:conf/icse/YangSH022, DBLP:journals/tosem/ZhouZSHCG22}.
For datasets, we not only use the CodeXGLUE benchmark studied in many original papers on PTMCs and adversarial attacks, but also include a new dataset, the OWASP benchmark, to evaluate the general applicability of the attack approaches.
For the target models, we mitigate this threat by selecting the most popular PTMCs with relatively high performance.

	\vspace{-2mm}
\section{Related work}
\textit{Black box attack} approaches have been extensively discussed in Section~\ref{sec:attack_method},  and we introduce other \textit{white box attack} approaches in this section.
Specifically, Yefet~\emph{et al.}~\cite{DBLP:journals/pacmpl/Yefet0Y20} propose DAMP, which utilizes the gradient information of the target model to find replacement identifiers in the opposite direction of the gradient descent.
Meanwhile, they use \textit{one-hot vector} to encode code, aiming at obtaining candidates by perturbing the vector and then mapping them back to tokens.
However, such an approach is  unable to constrain the candidates so that it may obtain irrelevant identifiers similar to random substitutions.
Srikant~\emph{et al.}~\cite{DBLP:conf/iclr/Srikant0MCFZO21} turn the adversarial attack into an optimization problem, and identify two aspects in the adversarial attack: which parts of the program to transform, and what transformations to use. They correspond to the search strategy and replacement strategy respectively as we mentioned above.
Then, they use \textit{projected gradient descent} (PGD) based \textit{joint optimization} (JO) solver to obtain the optimal transform location and transform method.
Zhang~\emph{et al.}~\cite{DBLP:journals/tosem/ZhangFLMZYSLJ22} propose CARROT, which incorporates gradient information into transform operations to guide the search process more effectively. 
Although retrieving gradients during transform operations may take more time, it can effectively reduce search iterations.
The above white box attacks are not very practical as the latest SOTA models, such as ChatGPT, are increasingly becoming closed-source. 
These models are typically deployed remotely and offer services through API interfaces, making it difficult to access their internal structure and parameters.

There is no comprehensive evaluation towards adversarial attack on PTMCs currently, and the study most similar to ours is that of Zeng~\emph{et al.}~\cite{DBLP:conf/issta/ZengTZLZZ22}. However, they mainly focus on evaluating the effectiveness of PTMCs while the adversarial attack approaches are not fully studied.
Specifically, they evaluate several attack approaches adapted from the field of \textit{natural language processing} (NLP), and focus on comparison in terms of ASR.
In contrast, our study specifically focuses on attacks designed for SE applications, and we have additionally evaluated attack efficiency and the quality of the generated adversarial examples. Moreover, the attack approach they proposed simply combines WIR and random replacement without incorporating the unique characteristics of programming languages and code intelligence tasks, which is a common shortfall for most existing studies on adversarial code attacks. In this work, for the first time, we generate adversarial examples by perturbing source code based on the contextual information of identifiers.

	\vspace{-2mm}
\section{Conclusion} 
This study thoroughly evaluates the performance, efficiency, and robustness of adversarial attacks on PTMCs. 
Results show that PTMCs are easily susceptible to adversarial perturbations, with varying levels of robustness among different tasks. The code summarization model is found to be the most vulnerable. Additionally, high-performing attack approaches often come with significant computational overhead.
The importance of different statements is also analyzed, revealing varying levels of sensitivity among different context identifiers to counterattacks.
Based on such findings, we propose a new approach, BeamAttack, which improves the effectiveness of attacks by 21.30\% and efficiency by 14.62\% compared to the existing approach ALERT using the same identifier substitution strategy.

\section{Data Availability}
The data, source code, and the results of this paper are available at: 

\noindent\url{https://github.com/CGCL-codes/Attack_PTMC}.
	\begin{acks}
We sincerely thank all anonymous reviewers for their valuable comments. 
This work was supported by the Key Program of Hubei under Grant No. 2023BAA024, the National Natural Science Foundation of China (Grant No. 62002125), and the Young Elite Scientists Sponsorship Program by CAST (Grant No. 2021QNRC001).
\end{acks}

    \balance
	\bibliographystyle{ACM-Reference-Format}
	\bibliography{bib/references}
	
\end{document}